%% file: main.tex
\documentclass[aps,physrev,reprint,superscriptaddress,amsmath,amssymb,longbibliography]{revtex4-2}
\input{common.tex}

\begin{document}
\input{sections/frontmatter.tex}
\input{sections/abstract.tex}
\maketitle
\raggedbottom
\input{sections/introduction.tex}
\input{sections/physical_constraints.tex}
\input{sections/feasible_set_visualization.tex}
\input{sections/lorenz_intervals.tex}

\input{sections/distinct_spectra.tex}
\input{sections/conductivity_information.tex}
\input{sections/scope_implications.tex}
\input{sections/conclusions.tex}
\input{sections/endmatter.tex}
\bibliography{refs}
\end{document}

\input{sections/frontmatter.tex}
\maketitle
\begin{center}\textbf{Supplemental Material}\end{center}
\input{sections/supplementary_body.tex}
\bibliography{refs}

%% file: common.tex
\usepackage{array,booktabs}
\usepackage{graphicx}
\usepackage[hidelinks]{hyperref}

%% file: sections/frontmatter.tex
\title{Limits of Lorenz number estimation from the temperature-dependent Seebeck coefficient}
\author{Naoki Sato}
\email[Contact author: ]{SATO.Naoki@nims.go.jp}
\affiliation{Research Center for Materials Nanoarchitectonics (MANA), National Institute for Materials Science (NIMS), Namiki 1-1, Tsukuba 305-0044, Japan}
\author{Takao Mori}
\affiliation{Research Center for Materials Nanoarchitectonics (MANA), National Institute for Materials Science (NIMS), Namiki 1-1, Tsukuba 305-0044, Japan}
\affiliation{Graduate School of Pure and Applied Sciences, University of Tsukuba, Tennodai 1-1-1, Tsukuba 305-8671, Japan}

%% file: sections/abstract.tex
\begin{abstract}
Separating electronic and lattice thermal conductivity in thermoelectric materials often relies on a Lorenz number estimated from the Seebeck coefficient using an assumed band and scattering model.
A central issue is how much information about the Lorenz number comes from the temperature dependence of the Seebeck coefficient, rather than from the assumed model.
Here, we formulate the minimum and maximum Lorenz numbers compatible with prescribed Seebeck coefficients as linear programming problems in a discrete spectral representation.
The analysis assumes known chemical potentials and an energy dependence that is common across the observation temperatures for each candidate spectrum.
The reference Seebeck coefficients are calculated from single-band and two-band models with carriers of the same sign.
Increasing the number of prescribed Seebeck coefficients from one to five reduces the Lorenz number interval width by approximately 97.6\% and 97.8\%, respectively, in the fixed finite spectral representation.
Despite this reduction, the single-band example retains an attained interval of approximately $[1.46453,\,2.37037]\times10^{-8}\,\mathrm{W\,\Omega\,K^{-2}}$ at 400 K.
Adding exact electrical conductivity values does not narrow these ranges when the spectral amplitude can be adjusted independently at each temperature.
Thus, even exact Seebeck coefficients and electrical conductivity values at all five temperatures need not uniquely determine the electronic thermal conductivity at 400 K.
By contrast, for the same single-band example, knowing the ratios of the spectral amplitudes across temperatures allows the electrical conductivity values to confine all compatible Lorenz numbers to an enclosing interval less than 1\% as wide as the five-temperature interval with independently adjustable spectral amplitudes.
These results show that temperature-dependent Seebeck coefficients strongly narrow the compatible Lorenz number range, while electrical conductivity can further constrain the spectral shape when the temperature dependence of the spectral amplitudes is independently restricted.
\end{abstract}

%% file: sections/introduction.tex
\section{Introduction}
\label{sec:introduction}
Controlling point defects, dislocations, nanoscale precipitates, and grain boundaries has been an effective strategy for reducing lattice thermal conductivity and improving thermoelectric performance~\cite{Biswas2012,Zhu2009,KimDislocations2015,Poudel2008,Liu2021}.
Assessing the contribution of these structural modifications requires distinguishing electronic and lattice contributions to the measured total thermal conductivity.
The Wiedemann--Franz law is commonly invoked to estimate electronic thermal conductivity from electrical conductivity through the relation $\kappa_e=L\sigma T$.
Here, $\kappa_e$ is the electronic thermal conductivity at zero electric current, $\sigma$ is the electrical conductivity, $T$ is temperature, and $L$ is the Lorenz number.
The familiar constant Sommerfeld value is a limiting result for degenerate carriers under appropriate scattering conditions, not a universal value for thermoelectric semiconductors~\cite{Wang2018,Putatunda2019}.
Estimating the electronic contribution therefore requires a Lorenz number appropriate to the material and temperature of interest.

Several approaches have been developed to estimate the Lorenz number from the Seebeck coefficient $S$ using specified models of band structure and carrier scattering~\cite{Kim2015,Kumar2024,KumarErratum2025}.
For a single parabolic band with acoustic phonon scattering, both quantities are functions of the same reduced chemical potential~\cite{Kim2015}.
Eliminating this common parameter gives a model-specific relation between the Seebeck coefficient and the Lorenz number.
Kim \textit{et al.}\ proposed a simple approximation to this relation~\cite{Kim2015}.
Kim \textit{et al.}\ also compared their approximation with Lorenz numbers obtained using more detailed band and scattering models for selected thermoelectric materials~\cite{Kim2015}.
An approximation for a single Kane band with acoustic phonon scattering instead uses $S$, the band gap $E_g$, and $T$ to estimate $L$~\cite{Kumar2024,KumarErratum2025}.

The reliability of such estimates depends on the electronic structure and carrier scattering mechanisms in the material.
The transport spectrum, expressed here as the spectral conductivity, describes the energy dependence of electrical conduction associated with electronic structure and carrier scattering~\cite{Wang2018,Putatunda2019}.
Nonparabolic dispersions, multiple bands, and energy-dependent scattering change the transport distribution and can alter the relation between the Seebeck coefficient and the Lorenz number~\cite{Wang2018,Thesberg2017}.
First-principles calculations by Putatunda and Singh found substantial deviations from simple estimates for PbTe and Mg$_3$Sb$_2$ under the scattering approximations studied, but better agreement for SrTiO$_3$ and ZrNiSn~\cite{Putatunda2019}.
These material-dependent deviations matter when lattice thermal conductivity is inferred by subtracting the estimated electronic contribution from the total thermal conductivity.
If the total thermal conductivity and the other thermal transport contributions are fixed, an error in $L$ changes the inferred lattice contribution.
At fixed electrical conductivity and temperature, using an estimated Lorenz number $L_{\rm est}$ instead of the actual effective value $L$ gives
\[
 \kappa_{\rm lat}^{\rm est}-\kappa_{\rm lat}=-\sigma T(L_{\rm est}-L).
\]
Overestimating $L$ therefore lowers the inferred lattice thermal conductivity by the same amount as it raises the estimated electronic contribution; the relative bias can be large when the lattice contribution is small.
Low-temperature high-field measurements on Bi$_{0.4}$Sb$_{1.6}$Te$_3$ illustrate how assumptions about $L$ can change the inferred effect of nanostructuring on lattice thermal conductivity~\cite{Candolfi2025}.

A remaining issue is how tightly a given temperature dependence of the Seebeck coefficient constrains the Lorenz number beyond a particular band and scattering model.
The relevant range is obtained by considering different spectral conductivities that reproduce the reference Seebeck coefficient at each temperature.
Within the common transport moment description used here, the Seebeck coefficient fixes a transport-weighted mean energy relative to the chemical potential, whereas the Lorenz number reflects the energy variance~\cite{Wang2018,MahanSofo1996}.
Changing temperature changes the Fermi window and couples several mean constraints through the same spectral shape.
We ask how widely $L$ can vary among spectra that reproduce the same prescribed temperature-dependent Seebeck coefficients.

Hirosawa \textit{et al.}\ reconstructed spectral conductivity and chemical potential from electrical conductivity and Seebeck coefficients, and used the reconstructed spectrum to evaluate electronic thermal conductivity~\cite{Hirosawa2022}.
Endo and Matsuura applied this reconstruction approach to electronic thermal conductivity in Al--Pd--Re quasicrystals~\cite{Endo2026}.
Hirosawa \textit{et al.}\ assumed a temperature-independent spectral conductivity, whereas we retain a common spectral shape and allow its amplitude to vary independently across temperatures~\cite{Hirosawa2022}.

Here, we determine Lorenz number bounds for Seebeck coefficients calculated at five temperatures from single-band and two-band reference models with carriers of the same sign.
The chemical potential is specified at each temperature, while candidate spectral conductivities are allowed to differ in energy dependence from the reference models.
Each candidate spectrum is nonnegative and represented on a finite set of energy points, with a common energy dependence across temperatures multiplied by a separately adjustable positive scale factor.
Using noise-free reference values and known chemical potentials isolates ambiguity in the spectral conductivity from observation errors and uncertainty in chemical potential.
We compare constraints at one, three, and five temperatures, establish attainable Lorenz number intervals, and examine their dependence on the finite spectral representation.
We also distinguish freely adjustable spectral amplitudes from known or partially constrained amplitudes when assessing the information supplied by electrical conductivity.
The temperature dependence strongly narrows the allowed Lorenz number range, while the remaining freedom in electronic thermal transport depends on which additional spectral constraints are imposed.

%% file: sections/physical_constraints.tex
\section{Seebeck coefficient constraints and the physical origin of nonuniqueness}
\label{sec:physics}
\subsection{Mean and variance of the energy distribution}
We use electron charge $-e$, with $e>0$, and a nonnegative spectral conductivity $\Xi(E,T)$ whose charge factor is already included.
We adopt a moment representation in which the electrical, thermoelectric, and electronic thermal coefficients are described by one common nonnegative spectrum, and define
\begin{equation}
 K_n(T)=\int_D \Xi(E,T)(E-\mu)^n
       \left(-\frac{\partial f}{\partial E}\right)dE,
 \label{eq:moments}
\end{equation}
where $f$ is the Fermi function, $\mu=\mu(T)$, and $D$ is the assumed energy support.
The convention is $\sigma=K_0$, $S=-K_1/(eTK_0)$, and
$\kappa_e=(K_2-K_1^2/K_0)/(e^2T)$ at zero electric current.
Thus the subtraction of the thermoelectric cross term is part of $L=\kappa_e/(\sigma T)$; omitting it would describe a different thermal response.
Energies are expressed in joules unless otherwise stated; displayed Seebeck coefficients in microvolts per kelvin are converted to volts per kelvin before use in equations.
The spectral conductivity $\Xi(E,T)$ is an energy-dependent transport function, whereas the electrical conductivity $\sigma(T)=K_0(T)$ is obtained by integrating it against the Fermi window.

For $K_0>0$, normalizing the Fermi-window-weighted spectral conductivity defines a probability density over energy.
This density and the reduced energy are
\begin{equation}
 p_T(E)=\frac{\Xi(E,T)(-\partial f/\partial E)}{K_0(T)},
 \qquad u=\frac{E-\mu(T)}{k_BT}.
 \label{eq:measure}
\end{equation}
Here $p_T(E)dE$ is the normalized transport weight over energy for a given candidate spectrum.
Writing $\langle a\rangle_T=\int_D a(E)p_T(E)dE$ gives
\begin{equation}
 S=-\frac{k_B}{e}\langle u\rangle_T,\qquad
 L=\left(\frac{k_B}{e}\right)^2
       \left(\langle u^2\rangle_T-\langle u\rangle_T^2\right).
 \label{eq:variance}
\end{equation}
These mean--variance identities follow the established transport distribution formulation~\cite{Wang2018,MahanSofo1996}.
The Seebeck coefficient fixes a mean energy relative to the chemical potential; the Lorenz number measures the variance around that mean.
Nonnegative weight and finite moments imply $L\geq0$.
In the ideal single-energy limit the variance vanishes.
The optimization of $ZT$ considered by Mahan and Sofo~\cite{MahanSofo1996} is a different problem from bounding $L$ at fixed multitemperature observations.
The temperature-dependent Fermi window makes the energy dependence of $p_T$ vary with temperature even when the underlying spectral shape is fixed.
For the negative Seebeck coefficients considered here, the transport-weighted mean energy lies above the chemical potential in the stated electron sign convention.

Multiplying the spectrum by a positive constant leaves both $S$ and $L$ unchanged.
Variation in $L$ therefore requires different spectral shapes whose normalized transport distributions satisfy the prescribed mean-energy constraints but have different energy variances at the target temperature.
At one temperature, a mean alone does not generally determine a variance.
At five temperatures, however, the mean constraints are coupled through the common spectrum.
All five distributions arise by applying their respective Fermi windows to the same nonnegative spectral shape.
We therefore determine how much the variance in Eq.~\eqref{eq:variance} can change among spectra satisfying all five mean constraints simultaneously.

\subsection{A common spectral shape constrained at five temperatures}
The reference observations are specified at 300, 350, 400, 450, and 500 K, with the Lorenz number evaluated at 400 K.
At these temperatures the chemical potentials $\mu_i$ are prescribed, and each candidate has one common nonnegative spectral shape with a positive temperature-dependent scale factor.
At each observation temperature $T_i$, let $\Xi(E,T_i)=A_i g(E)$, where $g\geq0$ is common and $A_i>0$ is otherwise unrestricted.
In the relaxation time approximation, a transport distribution weights electronic states by the squared velocity component along the transport direction and by the relaxation time~\cite{Wang2018,Putatunda2019}.
Here $g(E)$ specifies the common energy dependence of the spectral conductivity, and $A_i$ is a positive scale factor that multiplies the entire spectrum at temperature $T_i$.
These scale factors are also referred to as spectral amplitudes; known relative spectral amplitudes specify their ratios across temperatures.
The common spectral shape assumption excludes temperature-induced changes in that energy dependence, but retains the temperature dependence of the Fermi window and the prescribed $\mu(T)$.
In a finite representation, let $g_j$ be its nonnegative coordinates, $\epsilon_{ij}=E_j-\mu_i$, and $\beta_{ij}>0$ the zeroth moment coefficients with the spectral amplitude removed, including the quadrature weights and Fermi window.
The finite coefficient convention is specified in Sec. S2 A of the Supplemental Material~\cite{SupplementalMaterial}.
Then
\begin{equation}
 \begin{gathered}
 B_{n,i}[g]=\sum_j\beta_{ij}\epsilon_{ij}^{n}g_j,\\
 K_{n,i}=A_iB_{n,i}[g].
 \end{gathered}
 \label{eq:finite}
\end{equation}
The exact Seebeck coefficients $S_i^*$ impose
\begin{equation}
 \sum_j\beta_{ij}(\epsilon_{ij}+eT_iS_i^*)g_j=0.
 \label{eq:Srows}
\end{equation}
These are five simultaneous spectral shape constraints.
For any admissible nonzero spectral shape and any positive electrical conductivity values, the spectral amplitudes can be chosen independently at each temperature as $A_i=\sigma_i/B_{0,i}[g]$.
Thus independently adjustable spectral amplitudes allow the five electrical conductivity values to be matched without further restricting the spectral shape.
Positive zeroth moment coefficients ensure that every nonzero admissible shape has positive electrical conductivity, so the cross-multiplication and normalization below are well-defined.

At the target temperature $T_t=400$ K, define $x_j=\beta_{tj}g_j/B_{0,t}[g]$.
The $x_j$ are normalized transport weights at the target temperature: they are nonnegative and sum to unity.
This normalization removes only the irrelevant overall scale and converts Eq.~\eqref{eq:Srows} into $Cx=0$, where
\begin{equation}
 C_{ij}=\frac{\beta_{ij}}{\beta_{tj}}
   \left(\frac{\epsilon_{ij}}{k_BT_i}+\frac{eS_i^*}{k_B}\right).
 \label{eq:C}
\end{equation}
The allowed set is therefore the slice $\mathcal P=\{x\geq0:\,Cx=0,\ \boldsymbol1^{\mathsf T}x=1\}$ of the unit simplex.
It describes all normalized finite spectral shapes satisfying the prescribed Seebeck coefficients, without imposing proximity to a reference spectrum.
Because the target Seebeck coefficient is fixed exactly, the Lorenz number is affine in the normalized transport weights:
\begin{equation}
 H=\sum_j\left(\frac{\epsilon_{tj}}{k_BT_t}\right)^2x_j,
 \qquad L=\left(\frac{k_B}{e}\right)^2H-(S_t^*)^2
 \label{eq:linearL}
\end{equation}
Since the feasible set is defined by linear equalities and nonnegativity, the minimum and maximum of $L$ can be obtained by linear programming.
This is a finite moment-bounding problem over a convex admissible set~\cite{Bertsimas2005}.
The reference spectrum in the discrete representation satisfies the constraints, so at least one admissible vector exists.
The normalized feasible set is closed and bounded.
Since $L$ is affine in $x$, it attains a minimum and a maximum on this set.
We verify each reported endpoint by checking that its weights satisfy all constraints and that a matching dual bound rules out a better objective value.
The full equations and verification conditions are given in the Supplemental Material.
In this verification, the finite coefficients are treated as ratios of integers, so the defining equalities and inequalities are checked without a floating-point feasibility tolerance.
Exactness refers to the defined rational finite coefficients; the continuous interpretation is discussed in Sec.~\ref{sec:scope}.

%% file: sections/feasible_set_visualization.tex
\begin{figure*}[t]
 \centering\includegraphics[width=17cm]{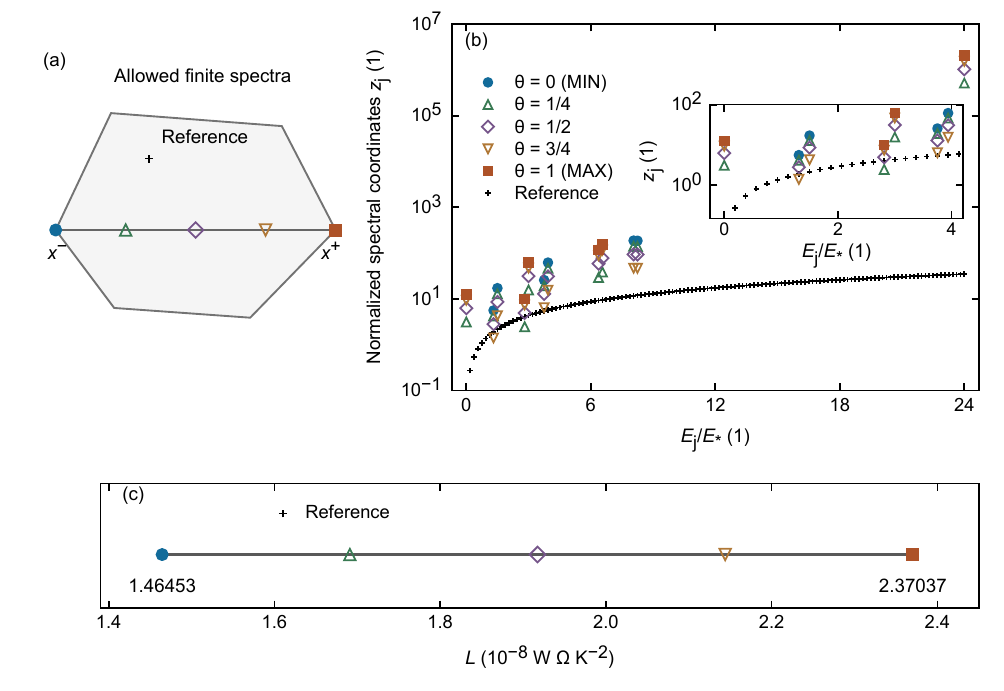}
 \caption{Admissible finite spectra and their Lorenz numbers for the original 129-node single-band (SB) example.
 (a) Schematic representation of $\mathcal P$, whose points represent complete spectral shapes satisfying the five Seebeck coefficient constraints and the target normalization.
 The outline is schematic rather than a calculated projection or section of the high-dimensional set.
 The five marked points on the line from $x^-$ to $x^+$ correspond to $x^{(\theta)}=(1-\theta)x^-+\theta x^+$ at $\theta=0,1/4,1/2,3/4,1$; the reference spectrum is shown separately.
 (b) Corresponding normalized spectral coordinates $z_j=x_j/\beta_{tj}=g_j/B_{0,t}[g]$ versus $E_j/E_*$, with $E_*=k_B(400\,\mathrm K)$.
 All candidates use the same target normalization.
 Only positive nodal coordinates are plotted on the logarithmic ordinate; omitted coordinates are exactly zero, and no interpolation between energy nodes is implied.
 The inset enlarges the low-energy region using the same coordinates and normalization.
 (c) The Lorenz numbers of the same candidates at 400 K.
 Marker colors and shapes identify the same candidates across panels; vertical marker offsets in (c), where used, are only for visual separation.
 The line from $L_{\min}$ to $L_{\max}$ is attained as $\theta$ varies continuously, although the displayed spectral family covers only part of $\mathcal P$.}
 \label{fig:feasible-set}
\end{figure*}

Figure~\ref{fig:feasible-set} relates the admissible spectral set to its Lorenz number interval.
In the schematic set shown in Fig.~\ref{fig:feasible-set}(a), each point represents an entire finite spectral shape after removal of its overall scale, rather than one energy point.
The constraints specify this set implicitly; its spectra need not be enumerated before the extrema of $L$ are determined.
For the original 129-node single-band (SB) example defined in Sec.~\ref{sec:intervals}, let $x^-$ and $x^+$ denote the minimizing and maximizing transport weights.
Their convex combinations satisfy
\begin{equation}
 \begin{aligned}
 x^{(\theta)}&=(1-\theta)x^-+\theta x^+,\\
 L\!\left(x^{(\theta)}\right)&=(1-\theta)L_{\min}+\theta L_{\max},
 \qquad 0\leq\theta\leq1.
 \end{aligned}
 \label{eq:convex-path}
\end{equation}
Every member of this family remains in $\mathcal P$ and reproduces the reference Seebeck coefficient at each of the five temperatures.
This path covers the complete Lorenz number interval shown in Fig.~\ref{fig:feasible-set}(c) but only part of the admissible spectral set in Fig.~\ref{fig:feasible-set}(a).

The optimization uses the normalized transport weights $x_j$.
To compare spectral shapes before Fermi-window weighting, Fig.~\ref{fig:feasible-set}(b) instead plots $z_j=x_j/\beta_{tj}=g_j/B_{0,t}[g]$ for five selected members of this family and the reference spectrum in the discrete representation.
Each spectrum is divided by its own target zeroth moment $B_{0,t}[g]$, so all displayed spectra satisfy $\sum_j\beta_{tj}z_j=1$.
The weights $x_j$ shown separately in supplementary Fig.~S2 include the target Fermi window and determine the mean and variance in Eq.~\eqref{eq:variance}.
The size of an unwindowed coordinate $z_j$ alone therefore does not determine its contribution to $L$; the corresponding transport weight and squared energy distance from the mean also matter.
The outline in Fig.~\ref{fig:feasible-set}(a) is schematic, whereas Fig.~\ref{fig:feasible-set}(b) and (c) use the original finite coefficients and endpoint solutions.

%% file: sections/lorenz_intervals.tex
\section{Lorenz number intervals for the two observation series}
\label{sec:intervals}
\subsection{Five-temperature reference series and attained intervals}
The reference Seebeck coefficients are calculated at 300, 350, 400, 450, and 500 K with nominal support $[0,24E_*]$, where $E_*=k_BT_*$ and $T_*=400$ K.
The SB spectral shape is $G_*(E/E_*)_+$, with $G_*=10^5\,\mathrm{S\,m^{-1}}$ and $(a)_+=\max(a,0)$.
Its linear energy dependence and nominal spectral amplitudes $A_i^0=T_*/T_i$ correspond to a three-dimensional single parabolic band with acoustic deformation-potential scattering in the elastic, equipartition approximation, with fixed band and scattering parameters~\cite{Wang2018,Putatunda2019}.
The TB spectral shape adds $G_*[(E-2E_*)/E_*]_+$ with equal relative coefficient and the same electron sign.

We prescribe $\mu(T)$ from a fixed-density single-band reference with $\mu(400\,\mathrm K)=0$.
Both the SB and TB reference series, and their respective inverse problems, use these same chemical potentials.
No separate two-band density condition is solved for TB.

The labels SB and TB identify the reference models that generate the prescribed Seebeck coefficients.
For each series, the inverse class with free spectral amplitudes is defined by the spectral constraints of Sec.~\ref{sec:physics}, without imposing the reference band count, linear spectral shape, or inverse-$T$ amplitudes.
The supplementary material specifies the nodes, coefficient rounding, density reference, and exact observations.

Figure~\ref{fig:intervals}(a) shows the two reference series of Seebeck coefficients.
The corresponding attained Lorenz number intervals are shown in Fig.~\ref{fig:intervals}(b).
\input{sections/figure_overview.tex}
For the original 129-node problems, the attained 400 K intervals are
\begin{equation}
 \begin{aligned}
 L_{\rm SB}&\in[1.46453,\,2.37037]\times10^{-8},\\
 L_{\rm TB}&\in[1.72260,\,2.63620]\times10^{-8}
 \end{aligned}
 \quad\mathrm{W\,\Omega\,K^{-2}}.
 \label{eq:results}
\end{equation}
The intervals are displayed with rounded endpoints; their defining values are exact rationals.
Every value between each pair is realized in its finite class, since convex combinations of the endpoint weights preserve all constraints and interpolate Eq.~\eqref{eq:linearL}.
For each reference series, the Seebeck coefficients calculated from the lower and upper endpoint spectra equal the same five prescribed values at their respective temperatures.
The SB and TB series themselves have different vectors of Seebeck coefficients and are two separate inverse problems.
These are complete compatibility ranges within the stated finite classes, rather than statistical confidence intervals.

The Lorenz numbers of the reference spectra evaluated in the discrete representation, shown in Fig.~\ref{fig:intervals}(b), lie inside their respective intervals, at approximately $1.6103\times10^{-8}$ for SB and $1.8716\times10^{-8}\,\mathrm{W\,\Omega\,K^{-2}}$ for TB.
Reference-model values obtained by continuous-energy integration and their evaluation conventions are given in the Supplemental Material.
The reference model supplies one spectrum and its Lorenz number within the wider class compatible with the five prescribed Seebeck coefficients.

For comparison with practice based on the Seebeck coefficient, we evaluate the approximation~\cite{Kim2015}
\begin{equation}
 \begin{split}
 L_{\rm Kim}={}&\left[1.5+\exp\!\left(-\frac{|S|}{116\times10^{-6}\,\mathrm{V\,K^{-1}}}\right)\right]\\
 &{}\times10^{-8}\,\mathrm{W\,\Omega\,K^{-2}}.
 \end{split}
 \label{eq:Kim}
\end{equation}
The inputs are the exact finite target observations, $S_{\rm SB}^*\simeq-204.71773$ and $S_{\rm TB}^*\simeq-227.25299\,\mu\mathrm{V\,K^{-1}}$.
They give $L_{\rm Kim}\simeq1.671\times10^{-8}$ and $1.641\times10^{-8}\,\mathrm{W\,\Omega\,K^{-2}}$, respectively, as shown in Fig.~\ref{fig:intervals}(b).
For SB, the difference between the approximate Lorenz number and the discrete reference value is much smaller than the width of the attained SB interval in Eq.~\eqref{eq:results}.
That interval spans Lorenz numbers from different spectral shapes that reproduce the same five prescribed Seebeck coefficients.
For TB, the estimate lies below the reported finite lower endpoint.
We obtained the TB lower endpoint using the prescribed $\mu(T)$ and all five exact Seebeck coefficients, whereas the formula of Kim \textit{et al.}\ assumes a single parabolic band and uses only the target $S$.
The TB reference spectrum therefore lies outside the approximation's single parabolic band assumptions.

The construction with carriers of the same sign has a simple physical interpretation.
For independent additive channels, let $\sigma_b$, $S_b$, and $L_b$ denote the electrical conductivity, Seebeck coefficient, and Lorenz number of channel $b$.
The weights are $w_b=\sigma_b/\sum_a\sigma_a$, and the total Seebeck coefficient is $S=\sum_bw_bS_b$.
Adding their moments before the open-circuit subtraction yields
\begin{equation}
 L=\sum_bw_bL_b+\sum_bw_b(S_b-S)^2.
 \label{eq:channels}
\end{equation}
The second term is the variance of the channel Seebeck coefficients weighted by electrical conductivity.
By Eq.~\eqref{eq:variance}, this variance reflects differences in the channel mean energies relative to the chemical potential.
A weighted average of the channel Lorenz numbers alone omits this term~\cite{Thesberg2017}.
It vanishes when the channel Seebeck coefficients coincide, but equal carrier signs do not impose that condition.
This explains why the Lorenz number of a two-band reference spectrum can differ from a single-band estimate without invoking bipolar carriers.
\input{sections/temperature_information.tex}

%% file: sections/figure_overview.tex
\begin{figure*}[t]
 \centering\includegraphics[width=17cm]{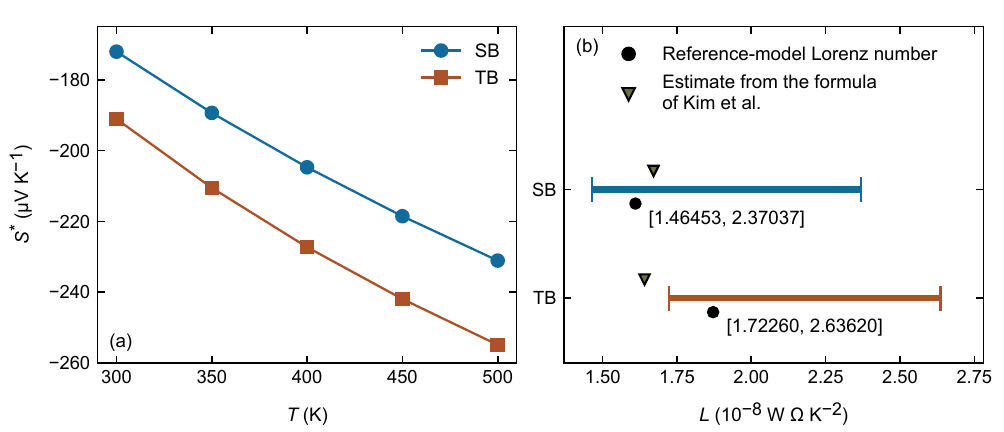}
 \caption{Prescribed Seebeck coefficients and attainable Lorenz number ranges.
 (a) Five exact finite Seebeck coefficients for the series generated by a single band (SB) and by two bands with carriers of the same sign (TB).
 Connecting lines guide the eye between the five observation temperatures.
 (b) Horizontal lines show the original 129-node attained intervals at 400 K.
 Filled circles mark the Lorenz numbers of the reference models evaluated in the discrete representation.
 Downward triangles mark estimates obtained by evaluating the approximate formula of Kim \textit{et al.}~\cite{Kim2015} [Eq.~\eqref{eq:Kim}] at each finite target Seebeck coefficient $S^*$.
 Vertical offsets of the markers from the interval lines are only for visual separation.
 Comparisons with continuous-energy reference evaluations are given in Sec.~S1~C of the Supplemental Material.
 Endpoint spectra within a series reproduce the reference Seebeck coefficient at each of the five temperatures; SB and TB are different series.
 For TB, this evaluation provides a comparison outside the single parabolic band assumptions underlying the approximation.}
 \label{fig:intervals}
\end{figure*}

%% file: sections/temperature_information.tex
\subsection{Temperature-subset comparison}
\label{sec:temperature-information}
We compare how strongly Seebeck coefficients at one, three, and five temperatures constrain the Lorenz number at 400 K.
Table~\ref{tab:temperature-information} gives four temperature choices separately for the SB and TB reference series.
Only the temperatures at which the reference Seebeck coefficients are imposed vary; the 129 energy points, finite coefficients, prescribed chemical potentials, energy range, and 400 K target remain fixed.
The chemical potentials are shared by the two series, but their Seebeck coefficients differ.
The chemical potential and energy range are prescribed inputs also in the one-temperature case.

\begin{table*}[t]
 \centering
 \caption{Attained Lorenz number intervals at 400 K for subsets of the reference Seebeck coefficients on the original 129 nodes.
 The minimum, maximum, and width $W$ are in $10^{-8}\,\mathrm{W\,\Omega\,K^{-2}}$.
 Width reduction is $100(1-W/W_1)$, where $W_1$ is the width for 400 K alone in the same series.
 Displayed values are rounded to nearest; widths are obtained from exact values before display rounding.}
 \label{tab:temperature-information}
 \input{sections/temperature_main_tabular.tex}
\end{table*}

Adding temperature information strongly constrains $L$: from one to five temperatures, the width decreases by approximately 97.6\% for SB and 97.8\% for TB.
Relative to the corresponding three-temperature interval, imposing all five observations further reduces the width by about 20.5\% for SB and 22.6--24.4\% for TB, depending on the three-temperature subset (Table~\ref{tab:temperature-information}).
These percentages quantify the contraction of the compatibility interval within the fixed finite model.
The five-temperature attained intervals in Eq.~\eqref{eq:results} nevertheless remain.
The derivation of the one-temperature bounds and the exact verification of the reported intervals are described in Sec. S3 C of the supplementary material.
The broad one-temperature intervals reflect the freedom to redistribute transport weights over the assumed energy range while preserving the fixed mean.
The endpoint values depend on both the assumed energy range and the locations of the discrete energy nodes.

The placement of the three temperatures also matters for these examples.
For both series, changing from 350/400/450 K to 300/400/500 K moves both endpoints downward while slightly increasing the interval width.
Because no inclusion relation is established between the two three-temperature feasible sets, this comparison is specific to the two temperature triplets considered here.

%% file: sections/temperature_main_tabular.tex
\begin{tabular}{llrrrr}\toprule
Series & \shortstack{Temperatures with prescribed\\Seebeck coefficients (K)} & Minimum & Maximum & Width & \shortstack{Width reduction\\(\%)}\\\midrule
SB & 400 & 0.00577 & 38.14797 & 38.14220 & 0.000\\
SB & 350, 400, 450 & 1.39971 & 2.53843 & 1.13872 & 97.015\\
SB & 300, 400, 500 & 1.27303 & 2.41235 & 1.13932 & 97.013\\
SB & 300, 350, 400, 450, 500 & 1.46453 & 2.37037 & 0.90584 & 97.625\\
TB & 400 & 0.00158 & 41.83516 & 41.83358 & 0.000\\
TB & 350, 400, 450 & 1.64537 & 2.82513 & 1.17976 & 97.180\\
TB & 300, 400, 500 & 1.47281 & 2.68145 & 1.20864 & 97.111\\
TB & 300, 350, 400, 450, 500 & 1.72260 & 2.63620 & 0.91359 & 97.816\\
\bottomrule
\end{tabular}

%% file: sections/distinct_spectra.tex
\section{Distinct spectra with nearly identical Seebeck coefficients}
\label{sec:profiles}
\begin{figure*}[t]
 \centering\includegraphics[width=17cm]{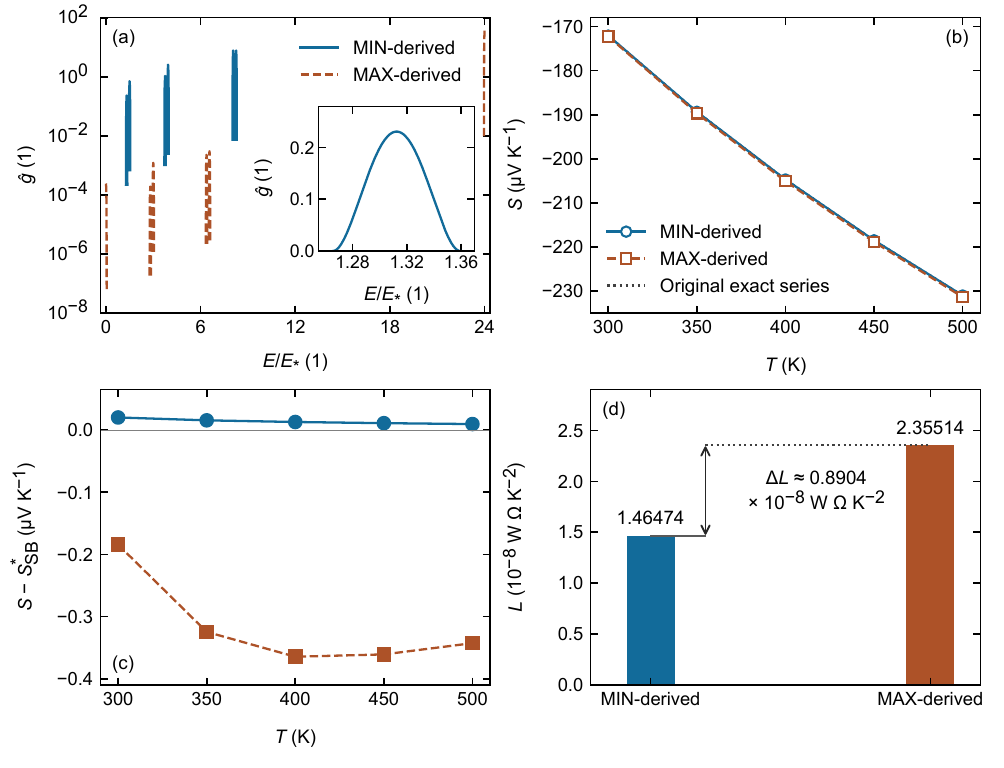}
 \caption{Finite-width spectra derived from the original SB endpoints.
 (a) The two profiles are shown over the full support using the same area-normalization convention, $\hat g(v)=E_*z_h(E_*v)/\sum_jm_j$, with $v=E/E_*$.
 A logarithmic ordinate makes small positive components visible.
 The inset enlarges the MIN component centered at $v=1.3125$, with half-width $h/E_*=0.046875$, using the same coordinates and normalization but a linear ordinate.
 (b) The original five-point series of Seebeck coefficients and the two finite-width responses nearly coincide on this scale; connecting lines are guides.
 (c) Their nonzero residuals are shown directly.
 (d) Bars show the Lorenz numbers of the two finite-width profiles at 400 K.
 The dotted guide and double-headed arrow indicate their difference, approximately $0.8904\times10^{-8}\,\mathrm{W\,\Omega\,K^{-2}}$.
 The responses are numerical evaluations of the finite-width profiles.
 The calculated Seebeck coefficients differ slightly from the prescribed values, so these profiles do not satisfy the original exact observation constraints.}
 \label{fig:profiles}
\end{figure*}

Figure~\ref{fig:feasible-set}(b) displays admissible spectra in the finite nodal representation.
We next examine how one specified finite-width replacement changes the Seebeck coefficients and Lorenz numbers of the original SB endpoint pair.
Each endpoint has six nonzero contributions, each of which is replaced by a nonnegative biweight profile, a quartic polynomial shape over a finite energy interval.
Figure~\ref{fig:profiles}(a) shows the resulting pair.
The basis-center parameter of each profile is fixed at its original energy.
Where the energy support truncates a profile, we adjust its normalization to preserve the integrated spectral weight $m_j$ defined in Eq.~(S14) of the Supplemental Material.
The half-width is $h/E_*=0.046875$, approximately 1.62 meV.
This illustration uses one prescribed width and one endpoint pair.

To compare spectral shapes rather than transport weights, we use the dimensionless display profile
\begin{equation}
 \hat g(v)=\frac{E_*z_h(E_*v)}{\sum_jm_j},\qquad v=E/E_*.
 \label{eq:profiledisplay}
\end{equation}
Here $z_h(E)$ is the sum of the six finite-width components for the chosen endpoint, and $m_j$ is the integrated spectral weight of component $j$, as defined in the Supplemental Material.
For display, $\hat g$ is normalized to unit area in $v$.
Figure~\ref{fig:feasible-set}(b) instead uses nodal coordinates $z_j=g_j/B_{0,t}[g]$ normalized by the target zeroth moment, so the two ordinates use different normalizations.
Transport moments are evaluated from the underlying $z_h$; the unit-area scaling is only the plotting convention.

The shapes shown in Fig.~\ref{fig:profiles}(a) are visibly different.
The minimum-derived profile concentrates its displayed intensity at lower energies, whereas the maximum-derived profile has a dominant contribution at the high-energy boundary and much smaller lower-energy components on this normalization.
The high-energy component is large in the spectral profile before multiplication by the Fermi window.
At the upper energy boundary, the Fermi window is strongly suppressed, so a large unwindowed component can coexist with a small target-normalized transport weight $x_j$.
We plot the entire energy range to include the high-energy component and use a logarithmic ordinate to show the much smaller components.
The inset in Fig.~\ref{fig:profiles}(a) resolves the finite width of a low-energy peak.

The resulting Lorenz numbers at 400 K, shown in Fig.~\ref{fig:profiles}(d), are approximately $1.46474\times10^{-8}$ and $2.35514\times10^{-8}\,\mathrm{W\,\Omega\,K^{-2}}$.
Their separation is $8.904\times10^{-9}\,\mathrm{W\,\Omega\,K^{-2}}$, or about 98.3\% of the width of the original SB five-temperature attained interval.
For the residuals shown in Fig.~\ref{fig:profiles}(c), the maximum absolute Seebeck coefficient residual over the five temperatures is $0.02043\,\mu\mathrm{V\,K^{-1}}$ for the minimum-derived profile and $0.36447\,\mu\mathrm{V\,K^{-1}}$ for the maximum-derived profile.
Both residuals are nonzero, so neither profile is presented as an exactly feasible point of the original problem.
This particular finite-width construction retains a large Lorenz number separation while its calculated Seebeck coefficients remain close to the original SB values at the five temperatures, as shown in Fig.~\ref{fig:profiles}(b).
Its boundary behavior, numerical integration, and comparison with the point-mass evaluation are specified in the supplementary material; the physical scope is discussed in Sec.~\ref{sec:scope}.

%% file: sections/conductivity_information.tex
\section{Electrical conductivity and spectral-amplitude constraints}
\label{sec:conductivity}
Prescribed electrical conductivity values can add constraints on the spectral shape when independent information restricts the temperature dependence of the spectral amplitudes.
\subsection{Fixed amplitude ratios}
\begin{figure*}[t]
 \centering\includegraphics[width=13.5cm]{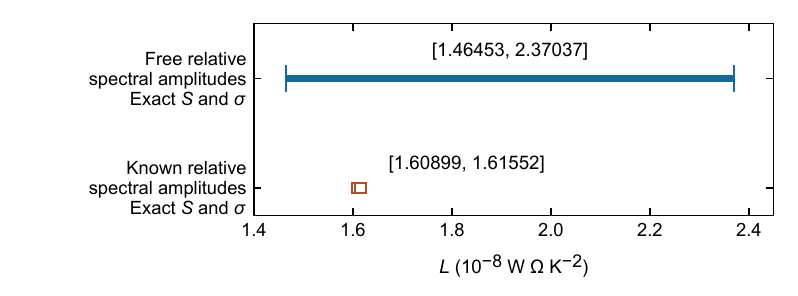}
 \caption{Effect of spectral amplitude information for the SB observations.
 Both conditions impose the same five exact Seebeck coefficients and positive electrical conductivity values (Table S5).
 The solid interval is attained with free relative spectral amplitudes, because independently adjustable positive spectral amplitudes can reproduce all five electrical conductivity values without restricting the shape.
 Open squares and a dashed connector show the certified outer enclosure with known relative spectral amplitudes.
 Attainability of the reported outer endpoints has not been established.}
 \label{fig:conductivity}
\end{figure*}

Figure~\ref{fig:conductivity} compares the attained SB interval for freely adjustable spectral amplitudes with the certified outer enclosure for amplitude ratios fixed to those specified by $A_i^0$.
Table S5 in the supplementary material lists these spectral amplitudes and the five exact electrical conductivity values, rounded for display.
Using the exact SB Seebeck coefficients and electrical conductivities gives the certified outer enclosure
\begin{equation}
 L\in[1.60899,\,1.61552]\times10^{-8}\,\mathrm{W\,\Omega\,K^{-2}}
 \label{eq:outer}
\end{equation}
at 400 K, with rounded values shown here.
Every Lorenz number compatible with these constraints lies within this interval.
Its sharpness and endpoint attainment have not been established.
For this SB reference series, its width is approximately 0.721\% of the width of the attained interval with free spectral amplitudes.
The comparison retains the same prescribed chemical potential and five exact Seebeck coefficients on the original 129 nodes.

With fixed amplitude ratios $A_i/A_t$, Eq.~\eqref{eq:finite} gives $B_{0,i}[g]/B_{0,t}[g]=(\sigma_i^*/\sigma_t^*)(A_t/A_i)$.
The prescribed electrical conductivity ratios therefore fix the relative zeroth moments.
The Seebeck coefficients fix $B_{1,i}[g]/B_{0,i}[g]=-eT_iS_i^*$, so the two sets of ratios constrain the common shape $g(E)$.
After target normalization, every such shape also satisfies the amplitude-free Seebeck coefficient constraints.

We obtain bounds on the dimensional second-moment quantity $\mathcal Q=K_{2,t}/(e^2T_t^2)$ and convert them into bounds on $L$.
For the same candidate with $\sigma_t=\sigma_t^*>0$, Eq.~\eqref{eq:linearL} gives $\mathcal Q/\sigma_t^*=(k_B/e)^2H$.
The quantity $\mathcal Q$ therefore retains the electrical conductivity scale removed by the normalization in $H$.
For known relative spectral amplitudes and fixed target observations,
\begin{equation}
 L=\frac{\mathcal Q}{\sigma_t^*}-(S_t^*)^2,\qquad \sigma_t^*>0.
 \label{eq:exacttarget}
\end{equation}
With $\sigma_t^*$ and $S_t^*$ fixed, only $\mathcal Q$ varies in Eq.~\eqref{eq:exacttarget}.
Bounds on $\mathcal Q$ therefore convert directly into bounds on $L$, without combining separate extrema of varying target quantities.
The conversion remains valid when observations at other temperatures have intervals while the target values remain fixed.

Keeping the amplitude ratios fixed, we also consider interval-valued Seebeck coefficients and electrical conductivities to represent observation uncertainty.
When target observations vary within intervals, combining separate bounds on $\mathcal Q$, $\sigma_t$, and $S_t^2$ can give an enclosing interval wider than the attainable Lorenz number range, because the separate extrema may occur for different spectra.
Sections S5 A and B of the supplementary material give the interval construction, the strengthening of the lower bound using $L\geq0$, and observation rounding conventions.
Comparing the outer-enclosure widths for interval-valued and exact observations does not isolate the effect of observation uncertainty.
Each width may exceed its corresponding attainable width by a different amount.

With fixed amplitude ratios, a common positive calibration factor can be accommodated by changing only the overall scale of $g$.
If $\sigma_{\rm obs}=c_{\rm cal}\sigma_{\rm true}$ at every temperature, replacing $g$ by $g/c_{\rm cal}$ reproduces the true electrical conductivities.
The common factor cancels from the moment ratios, leaving $S$ and $L$ unchanged, and the correspondence is reversible.
This requires that the allowed shapes remain allowed under positive global rescaling and that no independent absolute-scale restriction is imposed.
Independent temperature-dependent calibration factors generally cannot be accommodated by a single global rescaling.

\subsection{Bounded amplitude deviations}
Starting from the same five-temperature SB problem, we express partial spectral amplitude information as $A_i=A_i^0(1+\varepsilon_{A,i})$ with independent deviations $|\varepsilon_{A,i}|\leq\delta_A<1$ and $\delta_A\geq0$.
All five pairs of reference Seebeck coefficients $S_i^*$ and electrical conductivities $\sigma_i^*$ remain fixed exactly.
No physical normalization of $g$ or new absolute scale prior is imposed; $\delta_A=0$ recovers the known relative spectral amplitudes.
This problem retains the original 129 nodes, prescribed chemical potentials, and five exact SB Seebeck coefficient constraints, and adds restrictions on the amplitudes.
Every compatible Lorenz number therefore also belongs to the free-amplitude SB interval in Eq.~\eqref{eq:results}.
We intersect this interval with the moment-based outer enclosure to obtain the combined enclosing interval (supplementary Table S7).

For independent deviations bounded by $1\%$ at each temperature, it is approximately $[1.46453,\,1.78356]\times10^{-8}\,\mathrm{W\,\Omega\,K^{-2}}$.
This interval contains every compatible Lorenz number at 400 K and has approximately 35.2\% of the width of that free-amplitude interval.
The reference spectrum with zero amplitude deviations proves that the constrained set is nonempty; the combined outer endpoints have not been shown to be attained.
The $1\%$ bound applies independently to each amplitude; because all ratios share $A_t$, it is not an independent $\pm1\%$ bound on each $A_i/A_t$.

For a candidate shape $g$, the nominal electrical conductivity $\sigma_i^0[g]=A_i^0B_{0,i}[g]$ is the value calculated using the reference amplitude $A_i^0$.
The prescribed physical value satisfies $\sigma_i^*=(1+\varepsilon_{A,i})\sigma_i^0[g]$ for that same shape.
Equation~(S25) of the Supplemental Material expresses the allowed amplitude deviations as interval constraints on $\sigma_i^0[g]$, providing constraints on $g$ with the physical observations held fixed.

The corresponding nominal second-moment quantity is $\mathcal Q^0=A_t^0B_{2,t}[g]/(e^2T_t^2)$, evaluated with the reference target amplitude $A_t^0$.
The same factor $1+\varepsilon_{A,t}$ relates $\mathcal Q^0$ to $\mathcal Q_{\rm physical}$ and $\sigma_t^0[g]$ to $\sigma_t^*$, so it cancels in their ratio.
Thus $L=\mathcal Q^0/\sigma_t^0[g]-(S_t^*)^2$, as shown in Eq.~(S27) of the Supplemental Material.
Although the physical target electrical conductivity $\sigma_t^*$ is fixed, the nominal denominator $\sigma_t^0[g]$ varies with the allowed shape.
Bounds on $\mathcal Q^0$ must therefore be combined with the allowed nominal denominator rather than divided by the fixed physical value $\sigma_t^*$.
Combining separate bounds on this numerator and denominator can give an interval wider than the attainable Lorenz number range even with exact physical observations.
Section S5 C derives the equivalence and conversion in Eqs.~(S24)--(S27), with the affine conversion recovered at $\delta_A=0$.

%% file: sections/scope_implications.tex
\section{Discussion}
\label{sec:scope}
For the free-amplitude class, the attained interval in each specified finite representation consists of all Lorenz numbers obtained from nonnegative spectra that reproduce the prescribed Seebeck coefficients under the assumptions of Sec.~\ref{sec:physics}.
Each candidate has an energy dependence common across temperatures, with prescribed chemical potentials and independently adjustable positive amplitudes.
Adjusting those amplitudes also allows every admissible shape to reproduce the prescribed positive electrical conductivity at each temperature.
Consequently, even exact electrical observations permit different electronic thermal conductivities within this class.

To express this range in thermal-conductivity units, take $\sigma=10^5\,\mathrm{S\,m^{-1}}$ at 400 K as an illustrative value, independently of the reference-spectrum normalization and the electrical conductivities in Table S5.
Multiplying the original exact SB Lorenz-number endpoints by $\sigma T$ gives
\begin{equation}
 \kappa_e\in[0.58581,\,0.94815]\,
       \mathrm{W\,m^{-1}\,K^{-1}},
 \label{eq:thermal}
\end{equation}
with a separation of approximately $0.36234\,\mathrm{W\,m^{-1}\,K^{-1}}$.
Subtracting each endpoint estimate of electronic thermal conductivity from the same total thermal conductivity, after accounting for the same other thermal-transport contributions, gives two lattice thermal conductivity estimates separated by that same amount.
When the lattice contribution is small, this spread can be large relative to its inferred magnitude.
Such ambiguity can lead to a substantial relative error in an estimate based on a single assumed Lorenz number and can affect conclusions about the suppression of lattice thermal transport.

For the same-sign TB reference series, imposing all five Seebeck coefficients still leaves an attained Lorenz number interval whose width is comparable to that of the SB series on the original 129 nodes [Fig.~\ref{fig:intervals}(b) and Table~\ref{tab:temperature-information}].
The two series have different observation vectors, so this comparison demonstrates nonuniqueness for both examples without isolating the effect of band count.
For the TB series, the estimate from the single-parabolic-band formula of Kim \textit{et al.}~[Eq.~\eqref{eq:Kim}] falls below the finite lower endpoint.
That endpoint was obtained using the prescribed chemical potentials and all five Seebeck coefficients, whereas the formula uses only the target Seebeck coefficient.
The contrast illustrates the model dependence of this estimate for a multiband reference without bipolar transport.

We examined the finite representation by retaining every original coefficient, first adding midpoint nodes and then extending the upper energy limit.
The three representations have 129, 257, and 385 nodes, with nominal upper limits $24E_*$, $24E_*$, and $36E_*$.
The observations and chemical potentials are unchanged, and the node coordinates use the rational values defined in Sec. S2 of the Supplemental Material.
Any original solution remains feasible by leaving its original weights unchanged and setting all added weights to zero.
Its Lorenz number is unchanged, so the extended interval contains the original interval.
Replacing all quadrature weights on a finer grid would change the constraints.

From 129 to 385 nodes, the interval width increases by approximately 0.874\% for SB and 0.919\% for TB.
The 385-node endpoints are $[1.46304719,2.37680810]\times10^{-8}\,\mathrm{W\,\Omega\,K^{-2}}$ for SB and $[1.72223959,2.64422739]\times10^{-8}\,\mathrm{W\,\Omega\,K^{-2}}$ for TB.
Within each series, the lower endpoints at 257 and 385 nodes are exactly equal, while the upper endpoints increase.
The six comparisons are listed in Table S2 and illustrated in Fig.~S1(c) and (d) of the Supplemental Material.
Both original maximizing spectra assign nonzero weight to the upper-limit node, making the chosen energy range important.
Wang \textit{et al.}~\cite{Wang2018} also examined energy resolution and cutoffs through numerical integration tests.
Here, the additional nodes enlarge the allowed spectral class while the original coefficients remain fixed.

For each of the six finite representation problems, we verified with exact rational arithmetic that both endpoint spectra satisfy all constraints and that matching dual bounds exclude Lorenz numbers outside the reported interval.
These certificates establish optimality for the defined finite coefficients and observations.
Extending this guarantee to a continuous spectral class would require a justified connection to the finite representation, including rigorous error bounds for numerical integration and the numerically determined chemical potentials.
The reported finite extensions do not establish continuum convergence.

The endpoint spectra solve the finite optimization problem and have nonzero weights at only a few energy nodes.
The number of nonzero weights should not be interpreted as a material's band or peak count.
No material-specific restrictions on band dispersion, scattering, smoothness, or spectral intensity were imposed, and the physical realizability of the admissible spectra remains unassessed.
For the fixed finite-width profiles in Sec.~\ref{sec:profiles}, we calculated the Seebeck coefficients and Lorenz numbers.
The profiles retain a large Lorenz number separation.
Their calculated Seebeck coefficients differ slightly from the prescribed values, so the profiles do not satisfy the exact observation constraints.
Whether the two Seebeck responses could be distinguished within experimental measurement uncertainty remains to be assessed.

Within the present model class, electrical conductivity further constrains the spectral shape when the temperature dependence of the amplitudes is restricted, as in Sec.~\ref{sec:conductivity}.
Fitting a chosen band and scattering model to the same electrical observations leaves open alternative compatible spectra with different Lorenz numbers.
Material-specific restrictions therefore need support beyond that fit to justify a narrower Lorenz-number range.
Such restrictions could concern spectral widths, intensity ratios, or shapes derived from band structure or scattering.
We have not evaluated how these restrictions change the Lorenz number range or optimized $L$ over a material-specific class of smooth spectra.

%% file: sections/conclusions.tex
\section{Conclusions}
\label{sec:conclusions}
Temperature-dependent Seebeck coefficients strongly constrain the Lorenz number in the fixed finite spectral class studied here.
Increasing the prescribed observations from 400 K alone to five temperatures reduces the interval width by approximately 97.6\% for the single-band reference series and 97.8\% for the two-band series with carriers of the same sign.
Nevertheless, each series admits distinct spectra with the same five exact Seebeck coefficients and different Lorenz numbers, despite known chemical potentials and an energy dependence that is common across the observed temperatures for each candidate spectrum.
For the original single-band example, the attained interval is approximately $[1.46453,\,2.37037]\times10^{-8}\,\mathrm{W\,\Omega\,K^{-2}}$.
These spectra satisfy the same transport-weighted mean constraints but have different energy variances [Eq.~\eqref{eq:variance}].
The overall spectral scale cancels from the moment ratios.

The information supplied by electrical conductivity depends on the spectral amplitude assumptions.
With independently adjustable positive spectral amplitudes at the five temperatures, every admissible spectral shape can reproduce the prescribed positive electrical conductivity value at each temperature.
Exact Seebeck coefficients and electrical conductivity values at all five temperatures therefore still permit different electronic thermal conductivities at 400 K in this class.
For the same single-band observations and original finite representation, known ratios of the spectral amplitudes across temperatures allow the exact electrical conductivity values to confine all admissible Lorenz numbers to a certified outer enclosure less than 1\% as wide as the interval with free spectral amplitudes.
The narrowing reflects additional information about the temperature dependence of the spectral amplitude.
Partially constrained spectral amplitudes can also provide an additional restriction, as in the example with bounded deviations.
These outer enclosures remain distinct from proved attained intervals: their sharpness and endpoint attainability have not been established.

At fixed temperature and electrical conductivity, different admissible Lorenz numbers give different estimates of electronic thermal conductivity.
Subtracting a larger electronic estimate from the same total thermal conductivity gives a smaller residual by the same amount.
Interpreting that residual as lattice thermal conductivity requires the other thermal transport contributions to be accounted for consistently.
The reported numerical bounds are conditional on the specified finite representations and model-generated reference values.
In material-specific applications, the achievable constraint on $L$ will therefore depend on how much of the spectral shape and its temperature-dependent amplitude is known independently of the same Seebeck coefficient and electrical conductivity data.

%% file: sections/endmatter.tex
\begin{acknowledgments}
This work was supported by JST Mirai Program (JPMJMI19A1) and JSPS KAKENHI (JP26K17646).

The authors used ChatGPT and Codex (OpenAI) to assist with the development and review of computational codes and with manuscript preparation. The authors directed these activities, independently verified the AI-assisted outputs incorporated into this work, and take full responsibility for the reported results and manuscript.
\end{acknowledgments}

\smallskip
\noindent N. S. conceived the study, designed the research, developed the computational codes, performed the theoretical and numerical analyses, and wrote the original draft of the manuscript. N. S. and T. M. discussed the results and reviewed and edited the manuscript.

\section*{Data Availability}
The data that support the findings of this article are available from the authors upon reasonable request.

%% file: sections/supplementary_body.tex
\section{Conventions and the reference observations}
\subsection{Units and transport moments}
The carrier charge is $-e$, with $e=1.602176634\times10^{-19}$ C, and $k_B=1.380649\times10^{-23}\,\mathrm{J\,K^{-1}}$.
Both constants are represented by their exact SI decimal values.
The manuscript uses a spectral conductivity $\Xi$ with units $\mathrm{S\,m^{-1}}$; its charge factor is already included.
Consequently, $K_0=\sigma$, and $K_n$ has units $\mathrm{S\,m^{-1}\,J^n}$.
In an alternative convention with the charge factor removed, the spectrum and the electrical conductivity formula must be changed consistently.
The sign convention for the Seebeck coefficient is $S=-K_1/(eTK_0)$ and the Lorenz number is the open-circuit value.
All displayed $\mu\mathrm{V\,K^{-1}}$ values are converted to $\mathrm{V\,K^{-1}}$ before use in constraints.

Define the dimensionless window $w(u)=\exp(u)/(1+\exp(u))^2$.
Writing $u=(E-\mu)/(k_BT)$, the ideal Fermi window derivative is $w(u)/(k_BT)$.
The variance representation follows by normalizing $\Xi(-\partial f/\partial E)dE$ to unit mass.
It requires nonnegative spectral weight, positive electrical conductivity, and finite moments.
In the single-energy limit the variance vanishes; multiplication of the spectrum by a positive constant leaves $S$ and $L$ unchanged.
The channel-variance term in the main text follows from adding $K_n$ over independent channels before performing the open-circuit subtraction, consistent with the multiband transport analysis of Thesberg \textit{et al.}~\cite{Thesberg2017}.
It vanishes for identical channel Seebeck coefficients, but not generally for channels of the same carrier sign.

\subsection{Single-band reference and the prescribed chemical potential}
The nominal single-band spectral shape is $g(E)=G_* (E/E_*)_+$ with $G_*=10^5\,\mathrm{S\,m^{-1}}$ and $E_*=k_BT_*$, $T_*=400$ K.
Its transport exponent is one.
The nominal spectral amplitude $A_i^0=T_*/T_i$ is a positive factor multiplying the entire spectrum at temperature $T_i$.
The values $A_i^0$ specify the reference model.
In the amplitude-free inverse problem, the candidate amplitudes $A_i>0$ can be chosen independently at each temperature.
The target reduced chemical potential is $\eta_*=0$, with band edge at zero.
For the fixed-density parabolic band reference, define
\begin{equation}
 J_0^{(3/2)}(\eta)=\int_0^\infty u^{3/2}w(u-\eta)\,du,
 \qquad
 \left(\frac{T_i}{T_*}\right)^{3/2}
 \frac{J_0^{(3/2)}(\eta_i)}{J_0^{(3/2)}(0)}=1.
 \label{eq:si:density}
\end{equation}
The exponent $3/2$ in this density construction is distinct from the transport exponent one.
Integration by parts relates the integral in Eq.~\eqref{eq:si:density} to the usual parabolic band density integral with density of states proportional to the square root of energy.
The numerical roots, rather than exact roots of Eq.~\eqref{eq:si:density}, determine the chemical potential specified at each temperature.
The reference integration used absolute and relative tolerances $10^{-11}$, subdivision limit 300, and a dimensionless Fermi window width of 40.
The bisection stopping rule combined half-bracket width at most $10^{-10}$ (or a computed zero) with residual plus estimated error at most $10^{-9}$.
These numerical criteria are not rigorous root or integration enclosures.

For the exact finite model, write $\operatorname{RN}$ for binary64 rounding and $F(v)$ for the exact rational value of a stored binary64 number $v$.
The energy and chemical potential conventions are
\begin{equation}
 \begin{aligned}
 e_j^{\rm eV}&=\operatorname{RN}\!\left(\frac{24E_*j}{128e}\right),
 &E_j&=eF(e_j^{\rm eV}),\\
 \mu_i^{\rm eV}&=\operatorname{RN}\!\left(\frac{k_BT_iF(\eta_i)}{e}\right),
 &\mu_i&=eF(\mu_i^{\rm eV}),\\
 \epsilon_{ij}&=e\{F(e_j^{\rm eV})-F(\mu_i^{\rm eV})\},
 &A_i^0&=F\!\left(\operatorname{RN}(T_*/T_i)\right).
 \end{aligned}
 \label{eq:si:rounding}
\end{equation}
The notation $e_j^{\rm eV}$ denotes the numerical energy in electronvolts; multiplication by $e$ explicitly converts it to joules.
The SB reference spectrum in the discrete representation uses $g_j=G_* E_j/E_*$ on these nonnegative nodes.
For TB, the reference spectrum in the discrete representation samples the two linear contributions with threshold separation $2E_*$ and equal relative coefficient, while retaining the same $\mu_i$.
The exact finite Seebeck coefficients defined by moment ratios from those two reference spectra supply the two observation vectors in Table~\ref{tab:supplementary:1}.
Their displayed decimals are not inputs to the exact problems.

\begin{table}[!htbp]
 \centering
 \caption{Prescribed chemical potentials and exact finite observations, rounded for display.
 The chemical potentials are shared, but the SB and TB observation vectors differ.}
 \alttext{At each of the five temperatures, the prescribed chemical potential is common to the SB and TB series, but its value varies with temperature. The two observation series have different negative Seebeck coefficients, shown beside the common chemical potential at each temperature.}
 \label{tab:supplementary:1}
 \begin{tabular}{rrrr}\toprule
 $T$ (K) & $\mu$ ($10^{-21}$ J) & $S_{\rm SB}^*$ ($\mu$V/K) & $S_{\rm TB}^*$ ($\mu$V/K)\\\midrule
 300 & 2.3762101 & -172.07959 & -191.09163\\
 350 & 1.2511440 & -189.36157 & -210.51346\\
 400 & 0.0000000 & -204.71773 & -227.25299\\
 450 & -1.3646592 & -218.51803 & -241.87907\\
 500 & -2.8323589 & -231.03908 & -254.82343\\\bottomrule
 \end{tabular}
\end{table}

\subsection{Reference spectra in discrete and continuous evaluations}
For SB at 400 K, the reference spectrum evaluated in the discrete representation has $L=1.61026130\times10^{-8}\,\mathrm{W\,\Omega\,K^{-2}}$, compared with the continuous reference $1.61298366\times10^{-8}\,\mathrm{W\,\Omega\,K^{-2}}$.
For TB, the corresponding values are $1.8715667\times10^{-8}$ and $1.8747046\times10^{-8}\,\mathrm{W\,\Omega\,K^{-2}}$.
The TB target Seebeck coefficients are $-227.25299$ and $-227.05800\,\mu\mathrm{V\,K^{-1}}$, respectively.
These differences arise from the discrete and continuous evaluation conventions; the finite bounds use the discrete representation.
The original TB/SB interval-width ratio is approximately 1.00856 and compares two separate problems with different observation vectors.
The comparisons in the main text evaluate the approximation of Kim \textit{et al.}~\cite{Kim2015} at the two original finite target Seebeck coefficients.

\section{Finite coefficients and extensions retaining the original coefficients}
\subsection{Original 129-node coefficients}
Let $q_j^{\rm eV}>0$ be the trapezoidal weight and $q_{J,j}=eF(q_j^{\rm eV})$.
In ideal real arithmetic the original zeroth moment coefficient has the physical form
\begin{equation}
 \alpha_{ij}^{\rm ideal}
   =A_i^0\frac{q_{J,j}}{k_BT_i}
       w\!\left(\frac{E_j-\mu_i}{k_BT_i}\right).
 \label{eq:si:idealbeta}
\end{equation}
The zeroth moment coefficients are evaluated in binary64 arithmetic: the Fermi window in inverse electronvolts is multiplied first by the quadrature weight and then by the nominal amplitude.
The resulting binary64 value $a_{0,ij}$ is treated as the exact rational $\alpha_{ij}=F(a_{0,ij})$.
The zeroth moment coefficients with the spectral amplitude removed are exactly $\beta_{ij}=\alpha_{ij}/A_i^0$, using the rational amplitudes defined in Eq.~\eqref{eq:si:rounding}.
Because of intermediate rounding, $\beta_{ij}$ is defined by this evaluation convention rather than by an independent evaluation of the ideal expression in Eq.~\eqref{eq:si:idealbeta} divided by $A_i^0$.
The coefficients for the first and second moments are $\beta_{ij}\epsilon_{ij}$ and $\beta_{ij}\epsilon_{ij}^2$ in rational arithmetic.
Independently rounded values for the first or second moment are not substituted for them.

For a nonzero nonnegative finite shape, all five $B_{0,i}$ are positive since the defined $\beta_{ij}$ are positive.
The transformation $x_j=\beta_{tj}g_j/B_{0,t}[g]$ preserves all moment ratios.
Spectra that differ only by a positive global factor give the same normalized vector $x$.
Conversely, every normalized feasible vector determines a representative spectral shape through $z_j=x_j/\beta_{tj}$.
The normalization therefore removes the overall scale without imposing an additional restriction on the spectral shape.

\subsection{Enrichment and extension}
Index the original energy nodes by $j=0,\ldots,128$.
The refined energies satisfy
\begin{equation}
 E'_{2j}=E_j,\qquad E'_{2j+1}=(E_j+E_{j+1})/2,
 \label{eq:si:midpoint}
\end{equation}
where the odd-index equation applies for $j<128$.
The extended list retains $E'_0,\ldots,E'_{256}$ and appends
\begin{equation}
 E'_{256}+k\Delta,\qquad
 \Delta=(E_{128}-E_0)/256,\qquad k=1,\ldots,128.
 \label{eq:si:extension}
\end{equation}
All these operations use the exact rational energy values defined in Eq.~\eqref{eq:si:rounding}, without another rounding step.
The supports are nominally $[0,24E_*]$ and $[0,36E_*]$, but the actual coordinates retain their original binary64-derived rational values.

For an additional node only, the kernel convention is
\begin{equation}
 \begin{gathered}
 u_{ij}=\frac{E_j-\mu_i}{k_BT_i},\qquad \epsilon_{ij}=E_j-\mu_i,\\
 \beta_{ij}^{\rm new}=\frac{T_*}{T_i}
     F\!\left(w_{\rm binary64}(\operatorname{RN}(u_{ij}))\right).
 \end{gathered}
 \label{eq:si:newbeta}
\end{equation}
The ratio $u_{ij}$ is formed exactly, converted once to binary64, and passed to the dimensionless window implementation.
The positive window evaluated in binary64 is represented by its exact rational value; the first and second moment coefficients are then formed by exact multiplication with $\epsilon_{ij}$ and $\epsilon_{ij}^2$.
Here $T_*/T_i=E_*/(k_BT_i)$, where $E_*=k_BT_*$.
For an added-node coordinate $g_j$, Eq.~\eqref{eq:si:newbeta} therefore contributes $(E_*g_j)F(w_{\rm binary64}(\mathrm{RN}(u_{ij})))/(k_BT_i)$ to $B_{0,i}[g]$.
In this convention, $E_*g_j$ is the discrete spectral mass assigned to that added node.
The original columns are retained unchanged, including their canonical $C$ and $c$ entries.
SB and TB share the same kernel evaluations at additional nodes but have different constraint rows through their respective $S_i^*$.

If $I(j)$ maps an original node to its retained position, define $x'_{I(j)}=x_j$ and set all added coordinates to zero.
Retention of the original coefficients gives $A'x'=Ax=d$ and $(c')^{\mathsf T}x'=c^{\mathsf T}x$.
Thus every original feasible spectrum, including the reference spectrum and both original endpoints, remains feasible with the same Lorenz number within its series.
It also proves nesting after embedding the 257-node solutions in the 385-node problem.
If all additional weights are zero, the original objective is exactly retained; positive row or objective rescaling alone cannot change it.
The embedding enlarges the feasible set while preserving the original objectives, so the endpoint movements quantify the effect of the extended representation.

\section{Certificates, attainment, and endpoint weights}
\subsection{Canonical primal and dual problems}
For self-contained notation, let $B_{n,i}=\sum_j\beta_{ij}\epsilon_{ij}^ng_j$ and normalize by the positive target zeroth moment:
\begin{equation}
 x_j=\frac{\beta_{tj}g_j}{B_{0,t}},\qquad
 C_{ij}=\frac{\beta_{ij}}{\beta_{tj}}
 \left(\frac{\epsilon_{ij}}{k_BT_i}+\frac{eS_i^*}{k_B}\right),\qquad
 c_j=\left(\frac{\epsilon_{tj}}{k_BT_t}\right)^2.
 \label{eq:si:canonical}
\end{equation}
For the five-temperature problem, write $A=\begin{bmatrix}C\\\boldsymbol1^{\mathsf T}\end{bmatrix}$ and $d=(0,0,0,0,0,1)^{\mathsf T}$.
For $s=+1$ (minimum) or $s=-1$ (maximum), the signed primal and dual are
\begin{equation}
 \min_{x\geq0,\,Ax=d} sc^{\mathsf T}x,
 \qquad
 \max_{\pi\in\mathbb R^6,\,A^{\mathsf T}\pi\leq sc} d^{\mathsf T}\pi.
 \label{eq:si:primaldual}
\end{equation}
The objective gives $L=(k_B/e)^2c^{\mathsf T}x-(S_t^*)^2$.
An exact certificate consists of rational vectors $x,\pi$ satisfying
\begin{equation}
 Ax=d,\quad x\geq0,\quad sc-A^{\mathsf T}\pi\geq0,\quad
 sc^{\mathsf T}x=d^{\mathsf T}\pi.
 \label{eq:si:certificate}
\end{equation}
The dual multipliers have no sign restriction because the primal rows are equalities.
Weak duality follows directly from $x^{\mathsf T}(sc-A^{\mathsf T}\pi)\geq0$.
Thus feasibility and equality of the two objectives establish optimality for either sign.
Exact certification is based on the rational identities in Eq.~\eqref{eq:si:certificate}.

\subsection{Endpoint values and representation sensitivity}
\begin{figure}[!htbp]
 \centering\includegraphics[width=17cm]{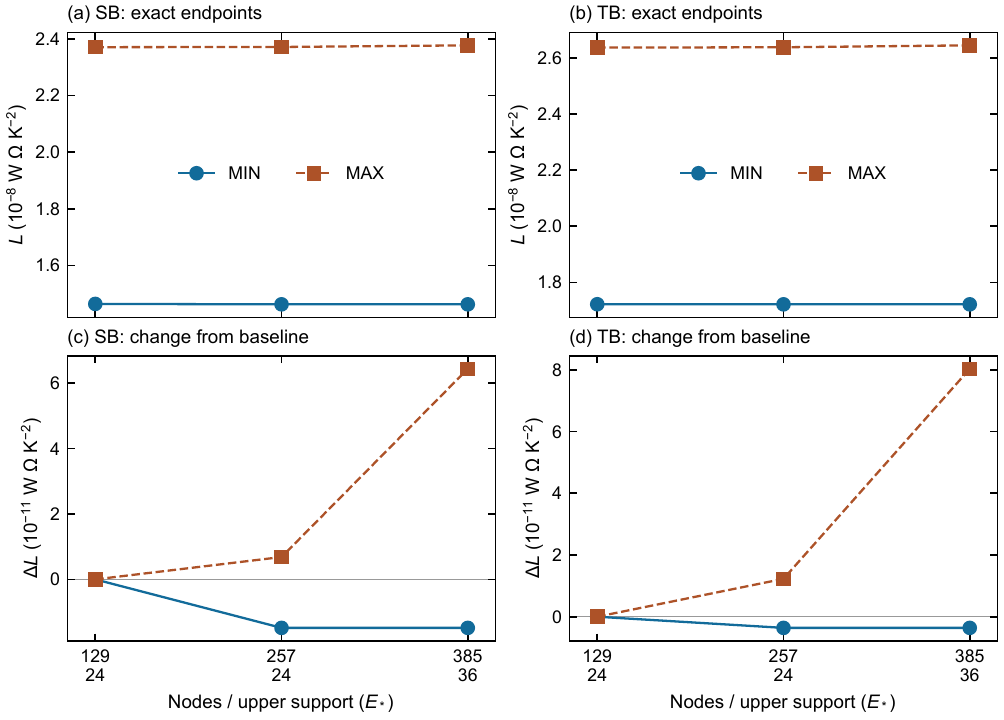}
 \caption{Exact Lorenz number endpoints under extensions retaining the original coefficients for (a) SB and (b) TB, with changes from the original 129-node values in (c) and (d).
 The three representations have 129, 257, and 385 nodes and nominal upper energy limits $24E_*$, $24E_*$, and $36E_*$.
 The figure label ``upper support'' denotes the upper energy limit; ``baseline'' denotes the original 129-node result.
 The observation vector is fixed within each series, and the original coefficients remain unchanged.
 Connecting lines identify the discrete comparison and do not establish continuum convergence.
 The two observation series have different constraints.}
 \alttext{The lower panels resolve small outward shifts of the lower and upper Lorenz number endpoints under the two extensions. All original feasible spectra remain allowed, so the extended intervals contain the original intervals.}
 \label{fig:si:representation}
\end{figure}

The endpoints for all six finite representations are listed in Table~\ref{tab:si:endpoints}.
Figure~\ref{fig:si:representation}(a) and (b) show the absolute endpoints for SB and TB, respectively; panels (c) and (d) show their changes under the two extensions retaining the original coefficients.
Within each series, every extended interval contains the original interval because each original feasible vector remains feasible when all added weights are set to zero.
The numerical comparison quantifies their expansion: the widths increase by approximately 0.874\% (SB) and 0.919\% (TB) from 129 to 385 nodes.
For each series, the exact minimum is unchanged between 257 and 385 nodes, whereas the maximum increases.

\begin{table}[!htbp]
 \centering
 \caption{Exact attained endpoint values and widths at 400 K, displayed in units of $10^{-8}\,\mathrm{W\,\Omega\,K^{-2}}$.
 Decimal entries are rounded for presentation; the certificates use rational values.
 Widths are calculated before display rounding.}
 \alttext{Six cases compare lower and upper Lorenz number endpoints and interval widths for two observation series on three node sets. The widths increase by approximately 0.874 percent for SB and 0.919 percent for TB between 129 and 385 nodes.}
 \label{tab:si:endpoints}
 \begin{tabular}{llrrr}\toprule
 Series & Nodes & Minimum & Maximum & Width\\\midrule
 SB & 129 & 1.46452974 & 2.37037116 & 0.90584142\\
 SB & 257 & 1.46304719 & 2.37105533 & 0.90800814\\
 SB & 385 & 1.46304719 & 2.37680810 & 0.91376091\\
 TB & 129 & 1.72260115 & 2.63619563 & 0.91359448\\
 TB & 257 & 1.72223959 & 2.63742311 & 0.91518352\\
 TB & 385 & 1.72223959 & 2.64422739 & 0.92198780\\\bottomrule
 \end{tabular}
\end{table}

Each of these six representation comparisons uses five rows for the Seebeck coefficients and one target normalization row.
For numerical conditioning, each row is divided by $\rho_i=\max(1,\max_j|A_{ij}|)$, and the objective by $\gamma=\max(1,\max_j|c_j|)$.
The normalization row has scale one.
Dual vectors are mapped back consistently before evaluating Eq.~\eqref{eq:si:certificate}.
Candidate solutions were obtained using \texttt{scipy.optimize.linprog} with the HiGHS method, presolve enabled, primal and dual feasibility tolerances $10^{-10}$, and interior-point optimality tolerance $10^{-12}$.
These floating-point solutions supply candidates whose final feasibility and optimality are verified by Eq.~\eqref{eq:si:certificate}.

For a recovered six-column basis of these five-temperature problems, the rational endpoint satisfies all six equalities, nonnegativity, and unit total weight; the submatrix formed by the six columns on the positive support has rank six.
The associated exact reduced costs have the required sign and the primal--dual gap is zero.
For either objective sign, weak duality and the primal--dual equality rule out any feasible vector with a better objective value.
Nonempty compactness guarantees the existence of extrema, while these certificates identify them.
If $x^-$ and $x^+$ are the two endpoints, $(1-\theta)x^-+\theta x^+$ is feasible for $0\leq\theta\leq1$ and gives the corresponding linear interpolation of $L$.
Thus an interval is attained, not merely two isolated values.

A residual-corrected outer bound is distinct from this exact optimum statement.
For any free-sign multiplier $\nu$ on scaled equality rows $\bar A x=\bar d$, set $r=sc+\bar A^{\mathsf T}\nu$.
Since $x\geq0$ and $\sum_jx_j=1$,
\begin{equation}
 sc^{\mathsf T}x\geq-\nu^{\mathsf T}\bar d
           +\min(0,\min_jr_j).
 \label{eq:si:equalityouter}
\end{equation}
This remains valid when some reduced costs are negative, but equality with an attained endpoint requires additional evidence.
If $r\geq0$, the residual correction vanishes and one recovers the usual dual bound.

\begin{table}[!htbp]
 \centering
 \caption{Positive support indices of the original exact endpoints, with zero-based node indexing.
 Both series have six active weights at each reported endpoint.}
 \alttext{Four lists identify the six positive node indices in each original endpoint. Minimum and maximum solutions use different nodes; both maximum solutions include the original upper support node numbered 128.}
 \label{tab:supplementary:2}
 \begin{tabular}{lll}\toprule
 Observation series & Endpoint & Positive node indices\\\midrule
 SB & MIN & 7, 8, 20, 21, 43, 44\\
 SB & MAX & 0, 15, 16, 34, 35, 128\\
 TB & MIN & 8, 9, 21, 22, 44, 45\\
 TB & MAX & 0, 16, 17, 35, 36, 128\\\bottomrule
 \end{tabular}
\end{table}

\begin{figure}[!htbp]
 \centering\includegraphics[width=17cm]{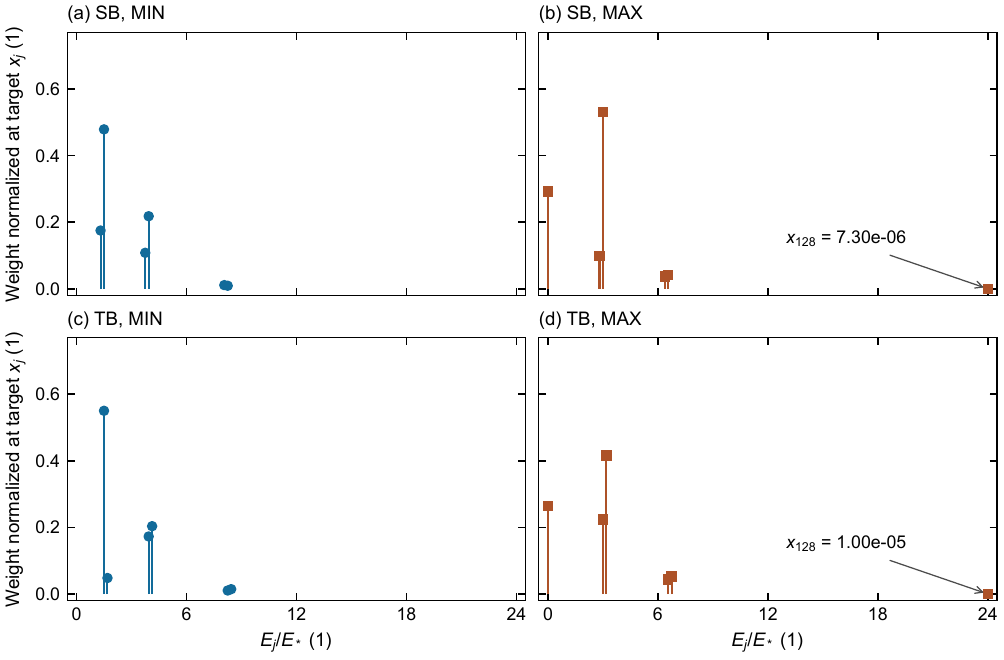}
 \caption{Exact weights normalized at the target temperature of the original 129-node endpoints, displayed as decimal coordinates.
 Only positive weights are plotted; all omitted weights are exactly zero.
 The ordinate shows the target-normalized transport weights $x_j$, with a separate stem for each positive weight, including those at closely spaced nodes.
 The six-node supports of these endpoint solutions arise from finite equality-constrained optimization and are not predictions of a material's peak structure.}
 \alttext{Four stem plots show normalized transport weights at the original minimum and maximum endpoints for the two observation series. Adjacent active nodes remain distinct, and arrows identify tiny nonzero weights at the upper boundary.}
 \label{fig:si:weights}
\end{figure}

The normalized weights are shown in Fig.~\ref{fig:si:weights}; they are distinct from the unwindowed spectral profiles.

\input{sections/temperature_certificates.tex}

\section{Finite-width construction and numerical evaluation}
For each original SB endpoint, its exact $x_j$ and the coefficients $\beta_{tj}$ and quadrature weights of the original discretization define
\begin{equation}
 z_j=\frac{x_j}{\beta_{tj}},\qquad
 q_{J,j}=eF(q_j^{\rm eV}),\qquad m_j=q_{J,j}z_j.
 \label{eq:si:masses}
\end{equation}
Here $z_j$ is dimensionless and $m_j$ has units of energy.
Only the six positive contributions are needed for each endpoint.
Let $D=[E_0,E_{128}]$, $h=(E_{128}-E_0)/512$, and
\begin{equation}
 \begin{split}
 a_j&=\max[-1,(E_0-E_j)/h],\qquad
 b_j=\min[1,(E_{128}-E_j)/h],\\
 P(t)&=t-\tfrac23t^3+\tfrac15t^5,\qquad N_j=P(b_j)-P(a_j),\\
 \psi_j(E)&=\frac{[1-((E-E_j)/h)^2]^2}{hN_j}
     \boldsymbol{1}_{\{E\in D,\ |E-E_j|\leq h\}},\qquad
 z_h(E)=\sum_jm_j\psi_j(E).
 \end{split}
 \label{eq:si:biweight}
\end{equation}
Each $\psi_j$ is nonnegative and integrates to one over $D$.
For an untruncated profile $N_j=16/15$; a profile centered exactly at a support edge has $N_j=8/15$.
Thus truncation conserves the prescribed mass by adjusting the normalization while keeping the basis-center parameter $E_j$ fixed.
It can produce a nonzero boundary value before extension by zero outside $D$, so global continuity or global $C^1$ smoothness is not claimed.
As $h$ tends to zero, these normalized measures formally approach their point masses for continuous test kernels; the reported calculation uses only the one fixed nonzero $h$.

We compare three evaluations: A denotes the original finite model, B evaluates the corresponding point masses with the continuous kernel, and C evaluates the finite-width profiles.
B uses $\sum_jm_j\delta_{E_j}$, and C uses $z_h(E)dE$.
At each temperature $T_i$, evaluations B and C use the same point-mass normalization.
Let $u_{ij}=(E_j-\mu_i)/(k_BT_i)$ and let $w_{ij}^{\rm pt}=w_{\rm binary64}(\mathrm{RN}(u_{ij}))>0$ be the point-window value evaluated in binary64.
Writing $u_{ij}(t)=u_{ij}+ht/(k_BT_i)$, the normalization and finite-width moments for each endpoint are
\begin{equation}
 \begin{aligned}
 \mathcal W_i&=\sum_j m_jF(w_{ij}^{\rm pt}),\qquad
 \omega_{ij}=\frac{m_jF(w_{ij}^{\rm pt})}{\mathcal W_i},\\
 J_n(T_i)&=\sum_j\omega_{ij}
   \int_{a_j}^{b_j}
   \frac{(1-t^2)^2}{N_j}
   \frac{w(u_{ij}(t))}{F(w_{ij}^{\rm pt})}
   [u_{ij}(t)]^n\,dt,\qquad n=0,1,2.
 \end{aligned}
 \label{eq:si:widthmoments}
\end{equation}
The sums run over the six nonzero basis contributions of the chosen endpoint.
The products, sums, and divisions defining $\mathcal W_i$ and $\omega_{ij}$ are performed in rational arithmetic before the normalized weights are converted to binary64.
Each integral is evaluated in the local coordinate $t$ over the truncated interval $[a_j,b_j]$.
At each quadrature node, the rational expression for $u_{ij}(t)$ is rounded to binary64 before evaluating the window; the remaining integrand arithmetic and weighted sums use binary64.
For evaluation B, the corresponding moments are $J_n^{(B)}(T_i)=\sum_j\omega_{ij}u_{ij}^n$.
The common factor $1/(k_BT_i)$ in the Fermi-window derivative cancels from the moment ratios at each temperature.
The finite-width zeroth moment is evaluated from Eq.~\eqref{eq:si:widthmoments} and retained in the ratios below.
Suppressing the temperature argument in $J_n$, the observables are
\begin{equation}
 S=-\frac{k_B}{e}\frac{J_1}{J_0},\qquad
 L=\left(\frac{k_B}{e}\right)^2
      \left[\frac{J_2}{J_0}-\left(\frac{J_1}{J_0}\right)^2\right].
 \label{eq:si:widthratios}
\end{equation}
The local integrations used absolute and relative tolerances $10^{-11}$ and subdivision limit 300.
The numerical integrator reports estimates of quadrature error.
These estimates are not rigorous error bounds for the Seebeck coefficient or Lorenz number calculated from Eq.~\eqref{eq:si:widthratios}.

\begin{figure}[!htbp]
 \centering\includegraphics[width=17cm]{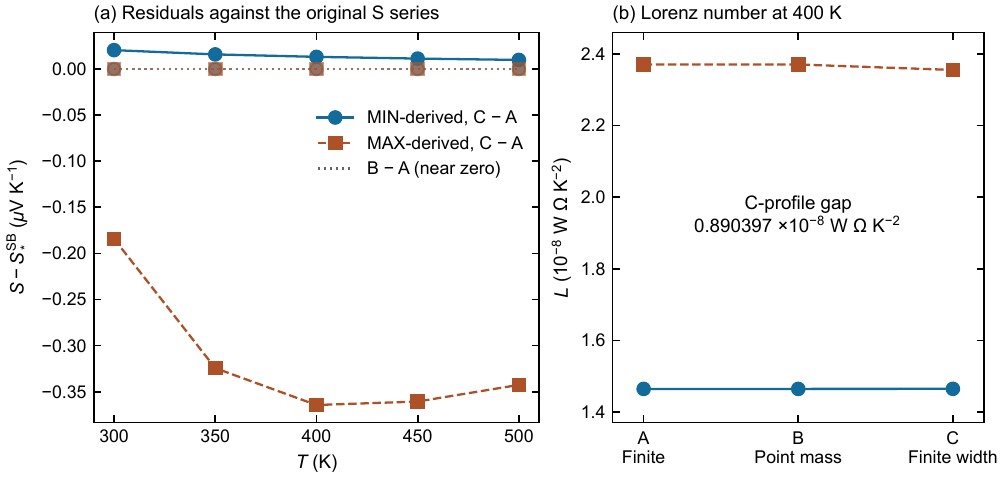}
 \caption{Finite (A), point-mass (B), and finite-width (C) evaluations for the original SB endpoint pair.
 (a) Seebeck coefficient residuals relative to the original exact series.
 The B--A residuals are near zero at the plotted scale, whereas the C--A residuals are nonzero.
 (b) Lorenz numbers at 400 K in the three evaluations.
 The C-profile gap is the Lorenz number difference between the two numerically evaluated finite-width profiles.
 Lines connect discrete evaluations only.}
 \alttext{Finite-width replacement causes small but nonzero Seebeck coefficient residuals while preserving most of the Lorenz number separation between the two endpoint-derived spectra. Point-mass evaluation remains close to the original finite evaluation.}
 \label{fig:si:ABC}
\end{figure}

At 400 K, B--A is zero in the binary64 Lorenz number comparison for MIN and $6.61744\times10^{-24}\,\mathrm{W\,\Omega\,K^{-2}}$ for MAX.
The C--B changes are $2.10335101\times10^{-12}$ and $-1.52337245\times10^{-10}\,\mathrm{W\,\Omega\,K^{-2}}$, respectively.
The C--A changes agree with these at the displayed precision, but the two differences are kept distinct in the underlying data.
Both C profiles have nonzero Seebeck coefficient residuals and are therefore numerical illustrations rather than feasible points of the exact-$S$ problem.
An internal consistency check reconstructed $J_0,J_1,J_2$ from the individual quadrature outputs of the six basis profiles, using the normalized point contributions as weights.
The reconstructed moments and the Seebeck coefficient and Lorenz number obtained from Eq.~\eqref{eq:si:widthratios} were compared with the corresponding moments and observables recorded for evaluation C.
Both routes use the same quadrature outputs, so this checks the weighted sums and the conversion to observables.
Figure~\ref{fig:si:ABC}(a) shows the Seebeck coefficient residuals, whereas Fig.~\ref{fig:si:ABC}(b) compares the 400 K Lorenz numbers in all three evaluations.

For the profile display in main text Fig.~3(a), $v=E/E_*$ and $\hat g(v)=E_*z_h(E_*v)/\sum_jm_j$.
Since each basis has unit mass on $D$, this convention gives unit area in $v$ for each whole profile.
It is the same global scale convention for MIN and MAX, not a new physical normalization or a rescaling of individual components.
The analytic biweight is sampled at 129 equally spaced local coordinates per basis, including each truncated endpoint, with the center also included; the union includes both support boundaries.
Zeros remain zeros in the data and are masked only on the logarithmic plot.
These display samples describe the prescribed polynomial profile and are distinct from the numerical integration used to evaluate transport moments.
The inset in main text Fig.~3(a) uses a subset of these coordinates around the MIN component centered at $v=1.3125$.
Its ordinate is linear; neither the profile nor its normalization is changed.
The curves of the Seebeck coefficient at five temperatures in main text Fig.~3(b) are numerical responses of the finite-width profiles, with residuals relative to the original exact observations shown in main text Fig.~3(c).

\section{Outer enclosures with electrical conductivity constraints}
\begin{table}[!htbp]
 \centering
 \caption{Model-generated SB reference inputs for the class with exact observations and electrical conductivity constraints in main text Fig.~4.
 Relative spectral amplitudes are fixed to $A_i^0$; their binary64-derived rational values follow Eq.~\eqref{eq:si:rounding}.
 The five electrical conductivity values are exact finite rational observations generated from the reference spectrum.
 All entries are rounded for display; the machine-readable data retain the exact values.}
 \alttext{Five temperatures from 300 to 500 kelvin list their nominal relative spectral amplitudes and exact model-generated reference electrical conductivity values rounded for display. The electrical conductivity decreases from about 101938 to 50845 siemens per meter.}
 \label{tab:si:sigma-inputs}
 \input{sections/sigma_inputs_tabular.tex}
\end{table}
Table~\ref{tab:si:sigma-inputs} gives the electrical conductivity values and relative spectral amplitudes corresponding to the SB Seebeck coefficients in Table~\ref{tab:supplementary:1}.
\subsection{Linear inequalities and ratio bounds}
This subsection and the observation-error analysis fix the relative spectral amplitudes to their nominal values $A_i^0$, unlike the amplitude-free class in the main results.
Let $\alpha_{ij}=A_i^0\beta_{ij}$ and write $K_{n,i}=\sum_j\alpha_{ij}\epsilon_{ij}^ng_j$.
For $0<\sigma_{i,-}\leq\sigma_{i,+}$ and $S_{i,-}\leq S_{i,+}$, admissibility imposes
\begin{equation}
 \begin{split}
 \sigma_{i,-}\leq K_{0,i}\leq\sigma_{i,+},\\
 -eT_iS_{i,+}K_{0,i}\leq K_{1,i}
                    \leq-eT_iS_{i,-}K_{0,i},\qquad g\geq0.
 \end{split}
 \label{eq:si:intervalconstraints}
\end{equation}
All these conditions are linear in $g$.
Because $K_{0,i}>0$, the bounds on the Seebeck coefficient can be rewritten as the linear inequalities in Eq.~\eqref{eq:si:intervalconstraints}.
When the observation intervals collapse, the paired inequalities become the corresponding exact equalities.

We define the dimensional second moment quantity
\begin{equation}
 \mathcal Q=\frac{K_{2,t}}{e^2T_t^2},\qquad
 [\mathcal Q]=\mathrm{S\,m^{-1}\,V^2\,K^{-2}}.
 \label{eq:si:dimensionalQ}
\end{equation}
This dimensional quantity is distinct from the dimensionless $H=c^{\mathsf T}x$ of the amplitude-free normalization.
Suppose certified linear bounds give $0\leq\mathcal Q_-\leq\mathcal Q\leq\mathcal Q_+$ over Eq.~\eqref{eq:si:intervalconstraints}.
Define $s_{\max}^2=\max(S_{t,-}^2,S_{t,+}^2)$ and let $s_{\min}^2=0$ if the target Seebeck coefficient interval contains zero, or $\min(S_{t,-}^2,S_{t,+}^2)$ otherwise.
Then a valid outer enclosure is
\begin{equation}
 \max\!\left(0,\frac{\mathcal Q_-}{\sigma_{t,+}}-s_{\max}^2\right)
 \ \leq L\leq
 \frac{\mathcal Q_+}{\sigma_{t,-}}-s_{\min}^2.
 \label{eq:si:outerL}
\end{equation}
The maximum with zero in Eq.~\eqref{eq:si:outerL} enforces $L\geq0$.
When the target observations vary within intervals, the extrema used for $\mathcal Q$, $\sigma_t$, and $S_t^2$ may correspond to different spectra.
Combining these separate extrema can give an interval wider than the attainable Lorenz number range even if the linear moment bounds themselves are sharp.

For exact target observations with $\sigma_t^*>0$, Eq.~\eqref{eq:si:outerL} specializes to
\begin{equation}
 \max\!\left(0,\frac{\mathcal Q_-}{\sigma_t^*}-(S_t^*)^2\right)
 \ \leq L\leq
 \frac{\mathcal Q_+}{\sigma_t^*}-(S_t^*)^2.
 \label{eq:si:exacttarget}
\end{equation}
In Eq.~\eqref{eq:si:exacttarget}, the target values $\sigma_t^*$ and $S_t^*$ are fixed, so only $\mathcal Q$ varies in $L=\mathcal Q/\sigma_t^*-(S_t^*)^2$.
This affine conversion does not require combining extrema of independently varying target quantities.
The same conclusion holds when observations at other temperatures have intervals while the target values remain fixed.
The exact-observation bounds in main text Fig.~4 are valid outer bounds, but their sharpness and endpoint attainability have not been established.

The outer bound below includes a correction for the dual residual.
The numerical coordinates and their connection to the dimensional second moment are
\begin{equation*}
 \begin{gathered}
 y_j=\frac{\alpha_{tj}g_j}{\sigma_{\rm ref}},\qquad
 \sum_jy_j=\frac{\sigma_t}{\sigma_{\rm ref}},\qquad
 d_+=\frac{\sigma_{t,+}}{\sigma_{\rm ref}},\\
 c_j=\left(\frac{\epsilon_{tj}}{k_BT_t}\right)^2,\qquad
 \mathcal Q=\sigma_{\rm ref}\left(\frac{k_B}{e}\right)^2c^{\mathsf T}y .
 \end{gathered}
\end{equation*}
Here $\sigma_{\rm ref}=10^5\,\mathrm{S\,m^{-1}}$ is a coordinate scale, not a constraint fixing the target electrical conductivity to that value.
For the exact observations in main text Fig.~4, $d_+\simeq0.6924141122$; unlike the amplitude-free weights $x$, these coordinates do not have unit sum.
In nonnegative scaled coordinates $y$ with $Ay\leq b$ and $\sum_jy_j\leq d_+$, take $\lambda\geq0$ and define the dual residual $r=sc+A^{\mathsf T}\lambda$.
The row imposing the upper bound on target electrical conductivity supplies $d_+$.
One has
\begin{equation}
 sc^{\mathsf T}y\geq-b^{\mathsf T}\lambda
       +d_+\min(0,\min_jr_j).
 \label{eq:si:inequalityouter}
\end{equation}
The correction in Eq.~\eqref{eq:si:inequalityouter} keeps the bound valid even when some components of $r$ are negative.
In Eq.~\eqref{eq:si:inequalityouter}, $A$ and $b$ include positive row scaling, while $c$ and $\lambda$ are expressed on the original objective scale.
The solver objective is $sc/\gamma$, with $\gamma=\max(1,\max_j c_j)$; its dual multipliers are rescaled by $\gamma$ before forming $\lambda$.
Applying this to both objective signs and multiplying by the positive dimensional scale gives $\mathcal Q_-$ and $\mathcal Q_+$.
This is an outer bound construction, not an attainment certificate.

\subsection{Observation-error structure}
The convention for relative errors in electrical conductivity is $\sigma_{{\rm obs},i}=(1+\epsilon_i)\sigma_{{\rm true},i}$, $|\epsilon_i|\leq\delta_\sigma<1$.
Solving for the positive true electrical conductivity gives the inverse interval
$[\sigma_{{\rm obs},i}/(1+\delta_\sigma),\sigma_{{\rm obs},i}/(1-\delta_\sigma)]$.
This is distinct from applying a forward relative perturbation to the observed center.
The centered cases use the exact finite reference values $S_i^*$ and $\sigma_i^*$ defined in Sec. S1 and Table~\ref{tab:si:sigma-inputs}, respectively, without additional rounding.
For a Seebeck coefficient half-width $\Delta S$ and relative electrical conductivity parameter $0\leq\delta_\sigma<1$,
\begin{equation}
 I_{S,i}=[S_i^*-\Delta S,S_i^*+\Delta S],\qquad
 I_{\sigma,i}=\left[\frac{\sigma_i^*}{1+\delta_\sigma},
                         \frac{\sigma_i^*}{1-\delta_\sigma}\right].
 \label{eq:si:centered}
\end{equation}
Zero values of either error parameter give the exact corresponding observation.
All centered interval endpoints are constructed in rational arithmetic, with no subsequent binary64 rounding.

The patterned case instead uses
$S_{{\rm obs},i}=\operatorname{RN}(S_i^*+\Delta S p_{S,i})$ and
$\sigma_{{\rm obs},i}=\operatorname{RN}(\sigma_i^*(1+\delta_\sigma p_{\sigma,i}))$.
The patterns are $p_S=(1/2,-1/2,0,1/2,-1/2)$ and $p_\sigma=(-1/2,0,1/2,-1/2,1/2)$.
The inverse intervals are formed from $F(S_{{\rm obs},i})$ and $F(\sigma_{{\rm obs},i})$ by the same additive and inverse-relative rules, without a second rounding of their endpoints.
The centered cases place the observation centers at the reference values, whereas the patterned case shifts those centers according to the specified patterns.

\begin{table}[!htbp]
 \centering\small
 \caption{Certified outer enclosures for the original SB class with electrical conductivity constraints at 400 K.
 The tabulated enclosures use the moment bounds above, without the additional intersection described below.
 Values are displayed in $10^{-8}\,\mathrm{W\,\Omega\,K^{-2}}$ and rounded for presentation, not outward decimal certification.
 The centered rows use Eq.~\eqref{eq:si:centered}; the Seebeck coefficient half-width where present is $2\,\mu\mathrm{V\,K^{-1}}$.
 These are not established attained intervals.}
 \alttext{Eight observation conditions compare certified outer Lorenz number bounds with fixed relative amplitudes. Columns give lower bounds, upper bounds, and enclosure widths. The common-calibration row has the same values as the exact-observation row.}
 \label{tab:supplementary:3}
 \begin{tabular}{lrrr}\toprule
 Observation condition & Lower & Upper & Width\\\midrule
 Exact $S$ and $\sigma$ & 1.60899 & 1.61552 & 0.00653\\
 Centered $S$ interval; exact $\sigma$ & 1.48091 & 1.72854 & 0.24763\\
 Exact $S$; centered $\sigma$, 1\% & 1.45300 & 1.78356 & 0.33055\\
 Exact $S$; centered $\sigma$, 2\% & 1.30483 & 1.94916 & 0.64433\\
 Exact $S$; centered $\sigma$, 5\% & 0.98048 & 2.46631 & 1.48583\\
 Both centered intervals, 5\% & 0.73369 & 2.70973 & 1.97604\\
 Both patterned intervals, 5\% & 0.68342 & 2.45286 & 1.76944\\
 Common positive calibration & 1.60899 & 1.61552 & 0.00653\\\bottomrule
 \end{tabular}
\end{table}

The last row of Table~\ref{tab:supplementary:3} assumes one positive calibration factor $c_{\rm cal}$ common to all five temperatures, with $\sigma_{\rm obs}=c_{\rm cal}\sigma_{\rm true}$.
If $g$ reproduces the observed electrical conductivities, $g/c_{\rm cal}$ reproduces the corresponding true electrical conductivities with the amplitude ratios held fixed.
The common factor cancels from the moment ratios, so $S$ and $L$ are unchanged.
This correspondence is reversible provided that positive global rescalings remain allowed and no independent absolute-scale constraint is imposed.
Independently varying calibration factors generally cannot be accommodated by one such rescaling.

For rows retaining all five exact SB Seebeck coefficients, the prescribed chemical potential, energy nodes, and base coefficients, every admissible shape also belongs, after normalization, to the amplitude-free class.
Let $I_L^{\rm free}=[L_{\min}^{\rm free},L_{\max}^{\rm free}]$ denote its attained interval and $I_L^{\rm outer}=[L_-^{\rm outer},L_+^{\rm outer}]$ the reported outer enclosure.
Their intersection supplies the combined outer enclosure
\begin{equation*}
 \begin{split}
 I_L^{\rm combined}
 &= I_L^{\rm outer}\cap I_L^{\rm free}\\
 &=\bigl[\max(L_-^{\rm outer},L_{\min}^{\rm free}),
         \min(L_+^{\rm outer},L_{\max}^{\rm free})\bigr].
 \end{split}
\end{equation*}
For the centered 5\% electrical conductivity intervals with exact $S$, intersection of the exact rational bounds equals $I_L^{\rm free}$, displayed approximately as $[1.46453,2.37037]\times10^{-8}\,\mathrm{W\,\Omega\,K^{-2}}$; the reference spectrum in the discrete representation ensures nonemptiness.
For the constrained problem, this intersection is an outer enclosure whose endpoint attainment has not been established.
The attainable range may still be narrower than $I_L^{\rm free}$.
The same $I_L^{\rm free}$ is not automatically applicable to rows with intervals for the Seebeck coefficient or with patterned observation centers.

\input{sections/amplitude_equivalence.tex}

\section{Numerical scope and reproducibility}
The exact results are defined by the rational observations and finite coefficient matrices together with endpoint vectors, dual multipliers, and the stated scaling conventions.
Verification of the exact identities requires these quantities; the nearest rounded values displayed in the text, tables, and figures are insufficient to reconstruct them and are not new mathematical inputs.
The representation-extension calculations used Python 3.14.4, NumPy 2.5.1, and SciPy 1.18.0 with single-thread settings.
The linear-program interface was SciPy's HiGHS method.

The exact certificates apply to the stated finite optimization problems.
The finite-width profiles are separate numerical illustrations evaluated by integration.

%% file: sections/temperature_certificates.tex
\subsection{Subsets of the observation temperatures}
\label{sec:si:temperature}
The temperature comparison isolates the information supplied by prescribing the Seebeck coefficient at additional temperatures within a fixed spectral representation.
Main text Table I compares eight temperature conditions on the original 129 nodes, distinct from the six representation comparisons in Table~\ref{tab:si:endpoints}.
The SB and TB series each use the subsets $\{400\}$, $\{350,400,450\}$, $\{300,400,500\}$, and $\{300,350,400,450,500\}$ K.
Only the temperatures at which Seebeck coefficients are prescribed vary; the energy points, finite coefficients, chemical potentials, energy range, target observation $S_t^*$, objective $c$, and target normalization remain unchanged within each series.
For a subset $I$ containing the target, let $C_I$ select the corresponding rows of $C$ and define
\begin{equation}
 \mathcal P_I=\{x\geq0:C_Ix=0,\ \boldsymbol1^{\mathsf T}x=1\},
 \qquad A_I=\begin{bmatrix}C_I\\\boldsymbol1^{\mathsf T}\end{bmatrix},
 \qquad d_I=(0,\ldots,0,1)^{\mathsf T}.
 \label{eq:si:subset}
\end{equation}
Equation~\eqref{eq:si:subset} has two, four, or six equality rows for one, three, or five temperatures, respectively.
The dual vector has the corresponding dimension; the six-column basis description above applies only to the five-temperature comparisons.
Imposing fewer Seebeck coefficient constraints gives $\mathcal P_5\subseteq\mathcal P_{3}\subseteq\mathcal P_1$ for either three-temperature choice, but supplies no mutual inclusion between the two three-temperature sets.
No inclusion between SB and TB is asserted.

At one temperature, put $u_j=\epsilon_{tj}/(k_BT_t)$ and $\bar u_t=-eS_t^*/k_B$.
The constraints become $\sum_jx_j=1$ and $\sum_ju_jx_j=\bar u_t$, with $x_j\geq0$.
For two distinct nodes $a<b$ that bracket the mean, weights
\begin{equation}
 x_a=\frac{b-\bar u_t}{b-a},\qquad
 x_b=\frac{\bar u_t-a}{b-a}
 \label{eq:si:two-node-weights}
\end{equation}
at those nodes, with all other weights zero, satisfy both equalities.
The subscripts $a,b$ in Eq.~\eqref{eq:si:two-node-weights} identify the nodes at those reduced energies.
If $a,b$ are adjacent nodes, the line $\ell(u)=(a+b)u-ab$ obeys $u_j^2-\ell(u_j)=(u_j-a)(u_j-b)\geq0$ on every allowed node.
It therefore bounds the second moment from below, and the two-node weights attain equality.
If instead $a=u_{\min}$ and $b=u_{\max}$, then $\ell(u_j)-u_j^2\geq0$ on the whole support, giving an attained upper bound.
Consequently,
\begin{equation}
 \begin{aligned}
 L_{\min}&=\left(\frac{k_B}{e}\right)^2(\bar u_t-a)(b-\bar u_t)
       &&\text{(adjacent bracketing nodes)},\\
 L_{\max}&=\left(\frac{k_B}{e}\right)^2
       (\bar u_t-u_{\min})(u_{\max}-\bar u_t).
 \end{aligned}
 \label{eq:si:one-temperature}
\end{equation}
When the mean equals a node, unit weight at that node gives $L_{\min}=0$, with the tangent $2\bar u_tu-\bar u_t^2$ providing the lower certificate.
When the mean equals either support endpoint, nonnegativity forces unit weight there and both extrema are zero; a mean outside the support is infeasible.
Equation~\eqref{eq:si:one-temperature} uses the actual finite rational nodes, not nominal endpoints substituted from the displayed support $[0,24E_*]$.
Its positive lower value between adjacent nodes is not a positive lower bound for arbitrary continuous spectra, which could put unit mass at the mean.

The one-temperature endpoints were constructed analytically.
For the two three-temperature conditions in both series, eight linear programs supplied candidate minima and maxima using the conditioning and solver tolerances described above, restricted to the selected rows.
Each candidate was converted to a rational four-column basis solution and a matching dual vector.
The five-temperature problems are identical to the original 129-node problems in Table~\ref{tab:si:endpoints}: they have the same coefficients, constraints, normalization, and objective.
The four endpoint vectors and dual certificates therefore establish the bounds for these same problems.
For all 16 endpoints, direct rational substitution checked Eq.~\eqref{eq:si:certificate} with $A_I,d_I$ and the appropriate dual dimension, including every reduced-cost inequality and an exactly zero primal--dual gap.
The nonempty sets $\mathcal P_I$ are compact simplex slices, and $L$ is affine because each subset includes the fixed target Seebeck coefficient.
Convex interpolation between the two certified endpoints therefore attains every intermediate Lorenz number.

Table~\ref{tab:si:temperature} gives the values underlying main text Table I at higher display precision.
Widths and reductions use exact rational values before rounding.
The denominator $W_1$ is the one-temperature width within the same series and fixed finite representation.
The wider three-temperature choice moves both extrema downward but leaves a slightly greater width in both series; this is not a general measurement-design result.
\begin{table}[!htbp]
 \centering
 \caption{Temperature-subset comparison on the original 129 nodes at the 400 K target.
 Minimum, maximum, and width are in $10^{-8}\,\mathrm{W\,\Omega\,K^{-2}}$; reduction is $100(1-W/W_1)$ in percent.
 All eight intervals are attained within their respective finite classes.
 Decimals are rounded to nearest, not outward bounds.}
 \alttext{Eight attained intervals compare the two reference series under one, two choices of three, and five prescribed observation temperatures. Five-temperature width reductions relative to the corresponding one-temperature width are 97.625094 percent for SB and 97.816122 percent for TB. The five-temperature cases impose the same constraints and objective as the original 129-node cases in Table S2.}
 \label{tab:si:temperature}
 \input{sections/temperature_sm_tabular.tex}
\end{table}

%% file: sections/temperature_sm_tabular.tex
\begin{tabular}{llrrrr}\toprule
Series & \shortstack{Temperatures with prescribed\\Seebeck coefficients (K)} & Minimum & Maximum & Width & \shortstack{Width reduction\\(\%)}\\\midrule
SB & 400 & 0.00577093 & 38.14796756 & 38.14219663 & 0.000000\\
SB & 350, 400, 450 & 1.39971037 & 2.53843454 & 1.13872417 & 97.014529\\
SB & 300, 400, 500 & 1.27303458 & 2.41235292 & 1.13931834 & 97.012971\\
SB & 300, 350, 400, 450, 500 & 1.46452974 & 2.37037116 & 0.90584142 & 97.625094\\
TB & 400 & 0.00158347 & 41.83516211 & 41.83357864 & 0.000000\\
TB & 350, 400, 450 & 1.64537269 & 2.82513421 & 1.17976152 & 97.179869\\
TB & 300, 400, 500 & 1.47280889 & 2.68145169 & 1.20864281 & 97.110831\\
TB & 300, 350, 400, 450, 500 & 1.72260115 & 2.63619563 & 0.91359448 & 97.816122\\
\bottomrule
\end{tabular}

%% file: sections/sigma_inputs_tabular.tex
\begin{tabular}{rrr}\toprule
$T_i$ (K) & $A_i^0$ (1) & $\sigma_i^*$ ($\mathrm{S\,m^{-1}}$)\\\midrule
300 & 1.3333333333 & 101938.0673\\
350 & 1.1428571429 & 83001.8773\\
400 & 1.0000000000 & 69241.4112\\
450 & 0.8888888889 & 58876.9316\\
500 & 0.8000000000 & 50844.7079\\
\bottomrule
\end{tabular}

%% file: sections/amplitude_equivalence.tex
\subsection{Exact observations with bounded spectral amplitude deviations}
\label{sec:si:amplitude}
All five pairs of SB reference Seebeck coefficients $S_i^*$ and electrical conductivities $\sigma_i^*$ remain fixed exactly, while independent deviations from the reference spectral amplitudes are allowed.
The spectral representation retains the original 129 energy points and a common nonzero nonnegative shape $g$.
With nominal positive amplitudes $A_i^0$, let
\begin{equation}
 A_i=A_i^0(1+\varepsilon_{A,i}),\qquad
 |\varepsilon_{A,i}|\leq\delta_A,\qquad 0\leq\delta_A<1,
 \qquad \sigma_i^0[g]=A_i^0B_{0,i}[g]>0.
 \label{eq:si:amplitude-box}
\end{equation}
The deviations $\varepsilon_{A,i}$ in Eq.~\eqref{eq:si:amplitude-box} are independent across temperatures.
Here $\sigma_i^0[g]$ is a nominal electrical conductivity, not the exact physical value $\sigma_i^*$.
The latter satisfies $\sigma_i^*=(1+\varepsilon_{A,i})\sigma_i^0[g]$ if and only if
\begin{equation}
 \frac{\sigma_i^*}{1+\delta_A}\leq\sigma_i^0[g]\leq
 \frac{\sigma_i^*}{1-\delta_A}.
 \label{eq:si:amplitude-inverse}
\end{equation}
The forward implication follows from positivity, and the inverse construction is
\begin{equation*}
 \varepsilon_{A,i}(g)=\frac{\sigma_i^*}{\sigma_i^0[g]}-1.
\end{equation*}
Any shape $g$ that satisfies both Eq.~\eqref{eq:si:amplitude-inverse} and the exact Seebeck coefficient constraints can reproduce the prescribed electrical conductivities by choosing the amplitude deviations given above.
Conversely, every allowed combination of a shape and amplitude deviations satisfies Eq.~\eqref{eq:si:amplitude-inverse}.
Positive spectral amplitudes cancel from the moment ratios, so the Seebeck coefficients and Lorenz numbers are unchanged.
The exact-observation problem with amplitude deviations and the problem with fixed nominal amplitudes and the conductivity intervals in Eq.~\eqref{eq:si:amplitude-inverse} therefore allow the same spectral shapes, the same normalized shapes, and the same Lorenz number values.
With the same relative bound, these are also the shapes allowed by the exact-Seebeck-coefficient, centered inverse-relative electrical conductivity problem in Sec. S5 B.
In that problem the intervals describe observation uncertainty, whereas here the observations are fixed and the spectral amplitudes are allowed to deviate.
No physical normalization of the global scale of $g$ or additional absolute scale prior is introduced.
At $\delta_A=0$, all deviations vanish and the known relative spectral amplitudes are recovered; the singular boundary $\delta_A=1$ is excluded.

For the second moment bounds, distinguish the nominal and physical quantities:
\begin{equation}
 \mathcal Q^0=\frac{A_t^0B_{2,t}[g]}{e^2T_t^2},\qquad
 \mathcal Q_{\rm physical}=(1+\varepsilon_{A,t})\mathcal Q^0,\qquad
 \sigma_t^*=(1+\varepsilon_{A,t})\sigma_t^0[g].
 \label{eq:si:amplitude-Q}
\end{equation}
Equation~\eqref{eq:si:amplitude-Q} gives
\begin{equation}
 L=\frac{\mathcal Q_{\rm physical}}{\sigma_t^*}-(S_t^*)^2
   =\frac{\mathcal Q^0}{\sigma_t^0[g]}-(S_t^*)^2.
 \label{eq:si:amplitude-L}
\end{equation}
In Eq.~\eqref{eq:si:amplitude-L}, bounds on $\mathcal Q^0$ cannot be divided directly by the fixed physical $\sigma_t^*$.
The nominal denominator varies within Eq.~\eqref{eq:si:amplitude-inverse}, so combining separate bounds on $\mathcal Q^0$ and $\sigma_t^0[g]$ can give an interval wider than the attainable Lorenz number range even though the physical $S_t^*$ and $\sigma_t^*$ are exact.
Only at $\delta_A=0$ does the nominal denominator collapse to $\sigma_t^*$, recovering the affine conversion in Eq.~\eqref{eq:si:exacttarget}.

Independent bounds on the amplitudes must also be distinguished from independent relative bounds on the amplitude ratios, because
\begin{equation*}
 \frac{A_i/A_t}{A_i^0/A_t^0}
 =\frac{1+\varepsilon_{A,i}}{1+\varepsilon_{A,t}}.
\end{equation*}
Because all ratios share the denominator $A_t$, independent bounds on the amplitudes do not correspond to independent relative bounds on $A_i/A_t$; the ratio for $i=t$ is identically one.

The equality of the admissible shape sets makes the four corresponding moment-based outer bounds in Table~\ref{tab:supplementary:3} valid for the amplitude conditions in Table~\ref{tab:si:amplitude}; the observation-error interpretation of the former table is unchanged.
The combined outer bounds are the intersection of those moment-based outer bounds with the five-temperature attained interval $I_L^{\rm free}$ for free spectral amplitudes.
The reference spectrum with zero deviations proves nonemptiness in all four cases.
At 5\%, the combined outer enclosure equals $I_L^{\rm free}$, but the amplitude-constrained attainable range may still be narrower because sharpness and endpoint attainment have not been established.
\begin{table}[!htbp]
 \centering
 \caption{SB outer enclosures for exact reference Seebeck coefficients and electrical conductivity values with independent bounded amplitude deviations.
 Bounds are in $10^{-8}\,\mathrm{W\,\Omega\,K^{-2}}$, rounded to nearest, not outward decimal bounds.
 The moment-based outer bounds use the separate nominal second-moment and electrical conductivity bounds; the combined outer bounds intersect them with the five-temperature attained interval for free spectral amplitudes.}
 \alttext{Four amplitude-deviation bounds of zero, one, two, and five percent give moment-based outer bounds and combined outer bounds obtained by intersection with the five-temperature attained free-amplitude interval. All four constrained sets are nonempty.}
 \label{tab:si:amplitude}
 \input{sections/amplitude_sm_tabular.tex}
\end{table}

%% file: sections/amplitude_sm_tabular.tex
\begin{tabular}{rrrrr}\toprule
$\delta_A$ (\%) & \multicolumn{2}{c}{Moment-based outer bounds} & \multicolumn{2}{c}{Combined outer bounds}\\
& Lower & Upper & Lower & Upper\\\midrule
0 & 1.60898945 & 1.61551998 & 1.60898945 & 1.61551998\\
1 & 1.45300145 & 1.78355627 & 1.46452974 & 1.78355627\\
2 & 1.30482686 & 1.94915610 & 1.46452974 & 1.94915610\\
5 & 0.98047608 & 2.46630605 & 1.46452974 & 2.37037116\\
\bottomrule
\end{tabular}